\documentclass[preprint,12pt,authoryear]{elsarticle}

\usepackage{amssymb}
\usepackage{amsmath}
\usepackage{url}

\usepackage{aas_macros}

\usepackage{color}

\usepackage{longtable}

\journal{New Astronomy Reviews}

\begin{document}

\begin{frontmatter}



\title{Natal kicks of compact objects} 


\author[Sternberg]{Sergei Popov}    
\author[Monash]{Bernhard M\"{u}ller}
\author[Monash,OzGrav]{Ilya Mandel} 

\affiliation[Sternberg]{organization={Sternberg Astronomical Institute, Lomonosov Moscow State University}, addressline={Universitetskij pr. 13}, city={Moscow}, postcode={119234}, country={Russia}}
\affiliation[Monash]{organization={School of Physics and Astronomy, Monash University}, city={Clayton}, state={Victoria}, postcode={3800}, country={Australia}}
\affiliation[OzGrav]{organization={OzGrav, Australian Research Council Centre of Excellence for Gravitational Wave Discovery}, country={Australia}}

\begin{abstract}
When compact objects -- neutron stars and black holes -- are formed in a supernova explosion, they may receive a high velocity at formation, which may reach or even exceed ${1000}$~km~s$^{-1}$ for neutron stars and hundreds of km~s$^{-1}$ for black holes. The origin of the velocity kick is intimately related to supernova physics. A better understanding of kick properties from astronomical observations will shed light on the unsolved problems of these explosions, such as the exact conditions leading to exotic electron capture and ultra-stripped supernovae. 

Kick velocities are profoundly important in several areas of astrophysics. 
Being a result of supernova explosions, the kick velocity distribution must be explained in the framework of the supernova mechanism.
The kick magnitudes and directions influence many topics related to binary systems, including the rate of compact object coalescences observable through gravitational waves. 
Moreover, knowledge of the kick velocity distribution is significant in predicting future observational results and their interpretation. For example, it is expected that the Roman space telescope will discover many microlensing events related to neutron stars and black holes; accurate estimates  of the number of observable microlensing events require precise kinematic properties of these compact objects.
\end{abstract}




\end{frontmatter}



\section{Introduction}

All the stars in our Milky Way Galaxy move with velocities that are related to their origin and age. The Milky Way stellar population is roughly divided into the halo, the bulge (with the bar), the thick disk, and the thin disk \citep{BinneyTremaine1987book, GilmoreReid1983MNRAS}. 
The oldest Galactic stellar population (Population II) consists of low-mass stars in the halo born with sub-solar metallicities  ($<10$\% of solar), as well as globular clusters and some stars in the bulge. The thick disk is sometimes called an intermediate Population II; it has an iron to hydrogen ratio of [Fe/H]$\lesssim-0.8$ \citep{2011ApJ...730....3S}. The halo has a roughly spherical shape with stars moving at random velocities comparable to or even exceeding the rotational velocity of the Milky Way. Due to a paucity of cold and dense gas,  star formation has ceased in the halo. 

The thin disk is very different from the halo. Its vertical height scale is $\approx 260$~pc, see \cite{Everall2022MNRAS} for more details on statistical and systematic errors. Most of the modern star formation in the Milky Way occurs in the thin disk and is mostly limited to the shining pattern of the spiral arms visible thanks to short-lived, luminous massive stars.  The thick disk is significantly older, composed of low-mass stars, and presently does not have significant star formation.  Thin and thick disks are frequently combined and jointly considered as the Milky Way disk. 

The vast majority of stars in the Milky Way disk follow relatively simple orbits. The stellar population of the  thin disk (Population I, roughly solar metallicity stars) rotates around the Galactic center in the same direction, following a roughly circular orbit with small excursions in radial and vertical directions.  Thus, a local standard of rest (LSR) is frequently introduced to describe the kinematic properties of the disk population. The LSR is the reference system that follows the exact circular orbit around the Galactic center at each radius. Individual stars have velocities that differ from this regular circular motion. This difference is called the peculiar velocity of a star. Most stars in the disk have small peculiar velocities, e.g. the Sun moves with 18~km~s$^{-1}$ with respect to its LSR \citep{Schonrich2010MNRAS}. 


Neutron stars (NSs) and black holes (BHs) are the remnants left behind by massive stars ($M \gtrsim 8$~M$_\odot$, see Section~\ref{s:models} for more details). 
 These OB stars have typical velocities $\sim10$~-~30~km~s$^{-1}$ relative to the LSR \citep{2022AstL...48..243B} and are mainly situated in the spiral arms of the Galaxy. 
Meanwhile, compact objects have very different kinematic properties and spatial distribution.
\footnote{The spatial distribution of NSs is also expected to be different from the distribution of their progenitors; the stationary distribution of kicked compact objects in the Galaxy was calculated semianalytically by \citet{1993A&AT....4...81P, 1994A&A...286..437P}.}  NSs are frequently found to move rapidly (hundreds of km~s$^{-1}$) with respect to the LSR or even with respect to their immediate birth environment (supernova remnant), see Section~\ref{s:direct_constraint}. Accreting X-ray binaries with NSs and BHs are found at large heights above the Galactic disk \citep{Repetto2012MNRAS,Repetto2017MNRAS}, see Section~\ref{s:indirect} for more details. 
Some isolated compact objects (especially, NSs) might exhibit kinematic behaviour more typical of Galactic Population II  as they also have elliptical orbits with large inclinations to the Galactic plane\footnote{It is still possible to distinguish the trajectories of objects kicked out from the Galactic disk from those of halo objects, such as those escaped from globular clusters, in the position/velocity or angular momentum/energy phase space.}, despite their progenitors clearly belonging to Population I. This apparent gap between the velocities of progenitors (massive stars) and their descendants (neutron stars and black holes) motivates the introduction of a natal kick: the velocity imparted when the compact object is born. 

 The large spatial velocities of isolated NSs play a crucial role in their long-term evolution \citep{2023Univ....9..273P}. 
Already in early 1970s, it was suggested that old isolated NSs might start to accrete from the interstellar medium \citep{1970ApL.....6..179O, 1971SvA....14..662S}.   
Such sources were much sought after in the 1990s \citep{1991A&A...241..107T} 
when the {\it ROSAT} X-ray observatory was operating. 
However, no sources of this kind were detected \citep{2000PASP..112..297T}. 
There are no reliable candidates for isolated accreting NSs so far. 
This may be a consequence not just of the low expected accretion luminosity of NSs, but may be due to evolutionary considerations. 
For large spatial velocities, isolated NSs are not expected to reach the accretion stage in a Hubble time, as the relativistic wind from the pulsar prevents accretion until the pulsar has significantly spun down \citep{2000ApJ...530..896P}. 
Only a fraction of NSs with the lowest kick velocities, below a few tens of km~s$^{-1}$, will have a sufficiently large Bondi radius (gravitational capture radius or accretion radius) to reach the evolutionary stage where accretion is possible late in their lives. Thus, precise knowledge about the velocity distribution of NS kicks is necessary to predict the observability of accreting isolated NSs, e.g., with the {\it eROSITA} survey \citep{2021ARep...65..615K}. 

Although the concept of a natal kick assumes that the entire velocity of the compact object arises in the first few seconds after core collapse or a supernova explosion, this need not always be the case.  For example, it was suggested that large peculiar velocities of NSs could be attributed to the so-called electromagnetic rocket \citep{HarrisonTademaru1975ApJ,AgalianouGourgouliatos2023MNRAS}. In this case, the NS is accelerated over time due to asymmetries in the pulsar operation. The magnetic field of the NS in this case is described as an off-center dipole that produces an asymmetric pulsar wind and accelerates the NS while the pulsar is active ($10-100$~Myr). In this case, the NS velocity immediately after the supernova explosion might be smaller than the velocity of the mature NS. Thus, the term natal kick is not entirely appropriate. Still, we use it below, keeping in mind the possibility of some contribution of an extended acceleration of an NS. It is worth noting that no analog of the electromagnetic rocket scenario has been proposed to explain BH kicks, since classical BHs do not experience mass loss and do not emit radiation on their own.  

High-velocity compact objects can also be born due to binary disruption 
\citep{1996ApJ...456..738I}, or dynamical interaction \citep{1967BOTT....4...86P}, or both \citep{2020ApJ...903...43D}. Binary disruption might happen directly at the moment of the birth of the compact object in a supernova (SN) explosion. The dynamical acceleration of the compact object can happen after its formation, or the progenitor can be accelerated before collapse, see e.g., \cite{2025A&A...701A..21D} and references therein.
In the latter case, the compact object progenitor is called a runaway star.  Runaway stars ejected due to dynamical interactions may have a velocity of up to $400~\mathrm{km\,s^{-1}}$ \citep{2009MNRAS.396..570G}. 
In the case of a binary disruption it is expected that the maximum ejection velocity of the progenitor is around $300\ \mathrm{km\,s^{-1}}$, or even up to $400\ \mathrm{km\,s^{-1}}$ \citep{2000ApJ...544..437P}, though see \citet{Renzo:2019}.

Some BHs, NSs, and binaries with compact objects are quite old and belong to Population II or were formed in globular clusters.  The kinematic properties of these compact objects and binaries are not necessarily attributed to large natal kicks. They rather reflect the properties of their formation site and further evolution in the changing Galactic gravitational potential. 
However, the number of these 
objects is expected to be quite limited because the stellar mass of Population II 
is less than 10~\% of Population I,  see e.g., \cite{2008A&ARv..15..145H, 2022MNRAS.514..689B, 2024ApJ...972..112C}, and references therein. 

  To summarize, a compact object can have a significant spatial velocity due to one or more of the following reasons:
 
 \begin{itemize}
     \item Natal kick {\it per se} via asymmetric neutrino emission, asymmetric jets and/or matter ejection, etc.;
     \item Gradual acceleration via asymmetric electromagnetic emission (only for NSs);
     \item Formation from a high-velocity progenitor, such as a runaway star from binary disruption.
 \end{itemize}

The Galaxy has other fast-moving stellar populations. Among the most interesting sources are hypervelocity 
white dwarfs \citep{Shen2018ApJ,Igoshev2023MNRAS}. 
For example, it is suggested that  the exploding white dwarf in a Type Iax supernova \citep{2017hsn..book..375J} can survive and is expelled with a large velocity of order hundreds of km~s$^{-1}$  \citep{2023MNRAS.518.6223I}. Also, the donor in a violent coalescence resulting in a Type Ia explosion can survive and obtain a spatial velocity $\sim 2000$~km~s$^{-1}$ \citep{2025arXiv251011781P}. Such events form only a small fraction of all Type Ia supernovae.
Naturally, a surviving white dwarf is not born in this event. Thus, the term natal kick cannot be properly applied in this case. 
 It is also theorized that white dwarfs might indeed receive a very weak natal kick of $\approx$0.75~km~s$^{-1}$  at formation \citep{ElBadry2018MNRAS}. This value was obtained by studying the separation distribution of wide binaries containing two main sequence stars and comparing it with the distribution of wide binaries containing a main sequence star and a white dwarf. The latter distribution has two breaks, which can be explained by white dwarf kicks \citep{ElBadry2018MNRAS}. We keep natal kicks of white dwarfs outside of our review since white dwarfs are not formed in supernova events and the estimated value of the kick barely affects binary dynamics. However, these kicks could still affect triple evolution, see e.g. \cite{Shariat2024arXiv}.

\section{Theoretical models of kicks from supernovae}
\label{s:models}

In this part of the review, we present various theoretical models for the origin of the kick. 
Section 2.1 is mostly devoted to potential mechanisms that can produce NS kicks and recent quantitative results from modern supernova simulations. BHs are discussed separately in Section 2.2. Many of the considerations on NS kicks also apply to BHs, so Section 2.2 focuses on the most salient differences in the BH case, backed up by a review of recent simulation results.

We open the section with a brief historical introduction.

\subsection{Neutron stars}

\subsubsection{Historical notes}

Historically, four main mechanisms of velocity gain by compact objects have been proposed. In chronological order these are: asymmetric matter ejection, i.e., hydrodynamical kicks \citep{1970SvA....13..562S}; binary disruption as an application of the \citet{1961BAN....15..265B} mechanism\footnote{See also \cite{1961BAN....15..291B}.} to NSs \citep{1970ApJ...160L..91G}; the electromagnetic rocket mechanism \citep{HarrisonTademaru1975ApJ}; and asymmetric neutrino emission \citep{1984SvAL...10...87C}.

\cite{1970SvA....13..562S} 
made the first proposal that NSs can attain large spatial velocities due to an asymmetric mass ejection during the supernova (SN) explosion. 
With very simple and naive estimates, he suggested that newborn compact objects can have velocities from a few hundred up to several thousand km~s$^{-1}$. Also, Shklovskii noted that with such high velocities, NSs can easily leave galaxies (at that time, dark matter halos were not yet known) and fill the intergalactic space. 

We want to underline that in his proposal, Shklovskii was not motivated by the necessity to explain large spatial velocities. In fact, his erroneous motivation was to explain pulsar spin-down by the Doppler effect and secular deceleration of a pulsar in the Galactic potential.  Nevertheless, he noted that the Crab pulsar, whose tangential velocity was measured to be 200 km~s$^{-1}$ at that time, requires a specific mechanism to explain this high value. 
The need for substantial birth velocities of pulsars was also noted by \citet{GunnOstriker1970ApJ}, though with less of a focus on explosion asymmetries as an explanation.
We will return to these asymmetries in the hydrodynamical kicks in section \ref{hydro}.

Meanwhile,
\cite{1970ApJ...160L..91G}  
aimed to explain the observed velocities of a few pulsars measured before 1970 and the width of pulsar distribution relative to the Galactic plane, see \cite{GunnOstriker1970ApJ}.
They applied the idea of the runaway stars proposed by \citet{1961BAN....15..265B}\footnote{See also \cite{1957moas.book.....Z}.} to the case of radio pulsars born in disrupted binary systems.

The authors considered a massive binary with semimajor axis $a$ in which both stars can produce pulsars and assumed that the system is  disrupted already after the first SN. Thus, it is assumed that $M_1-M_2 > 2 M_\mathrm{N1}$, so the system is  disrupted without a kick just due to an instantaneous mass loss equal to $M_1-M_\mathrm{N1}$. After the binary disruption, the newborn NS with mass $M_\mathrm{N1}$ born from the more massive companion with mass $M_1$ and the secondary component with mass $M_2$ have a relative velocity (at infinity)
\begin{equation}
    v_\infty= \left[ \frac{G(M_1-M_2-2M_\mathrm{N1})}{a}\right]^{1/2}.
\end{equation}
For realistic masses and $a \lesssim 0.1$~AU, it is easy to obtain $v_\infty \sim 100$~km~s$^{-1}$.   In this scenario, velocities up to a few hundred km~s$^{-1}$ can be achieved. This was sufficient to explain the data that existed at that time. 

Incidentally, \cite{GunnOstriker1970ApJ} already proposed a `Blaauw-like' explanation of pulsar velocities. They theorized that after the binary disruption, a more massive runaway star could serve as a `tug boat' for the newborn NS, allowing the latter to attain a velocity of $\sim 100$ km~s$^{-1}$. 

By 1975, the fact that pulsars had velocities about an order of magnitude larger than their progenitors was well-established. 
\cite{HarrisonTademaru1975ApJ} 
in their study addressed the origin of these velocities in the range $\sim 10-500$ km~s$^{-1}$.

They considered the acceleration of an NS by the electromagnetic radiation of a displaced dipole. In this case, the period-averaged reaction force acting on the NS is $\propto s\,\omega^5 $, where $s$ is the displacement (distance from the spin axis). Due to this strong dependence on the rotation frequency $\omega$, the acceleration rapidly vanishes as the pulsar spins down. Thus, the velocity is obtained mostly during the first year of the pulsar's life. If the pulsar is a member of a binary system, then on the same time scale, the orbit of the binary is modified. 
The scenario was developed and extensively applied to the known (at that time) population of pulsars by
\cite{1977ApJ...216..842H}. 
Modern calculations within the same framework can be found in
\cite{AgalianouGourgouliatos2023MNRAS},
 whose refined model tends to predict somewhat lower kicks than \citet{1977ApJ...216..842H}. 
These authors mainly apply the mechanism to model properties of radio pulsars in SNRs and conclude that the offset of the dipole might be huge, $\sim 7$~km,  and the initial spin period very small, $\sim 3-4$~ms, to explain the observed spatial velocities of these pulsars. 
Such rocket-type mechanisms are interesting for explaining the phenomenology of observed NSs in several respects.
Observations \citep{2007MNRAS.381.1625J,2012MNRAS.423.2736N,2013MNRAS.430.2281N,2024arXiv241212017B} suggest a tendency for spin-kick alignment among young NSs. An electromagnetic rocket
naturally leads to spin-kick alignment as
the period-averaged acceleration of the NS is directed
along the spin axis.

Recently, the electromagnetic rocket mechanism was criticized by \cite{2025arXiv250705602B}. 
These authors note that the assumption of the vacuum magnetosphere is not valid in a realistic situation.
In the framework of the force-free electrodynamics this kick mechanism is inefficient. 
\cite{2025arXiv250705602B} suggest that the rocket mechanism works in the case of a realistic magnetosphere only if pair production near the light
cylinder is suppressed. This condition is not satisfied for normal young radio pulsars and magnetars.

However, the electromagnetic rocket mechanism has several attractive features that may help explain elements of the phenomenology of NS spins and kicks that are difficult to account for by other mechanisms, such as the hydrodynamic mechanism discussed below. 
As the rocket mechanism does not immediately impart a kick at the time of explosion, but provides acceleration on longer time scales, its impact on NSs formed in  surviving  binaries  has additional features which influence the orbital parameters on a long timescale. 
Combining rocket-type kicks with acceleration acting over timescales much longer than the orbit with instantaneous hydrodynamic kicks at the time of explosion results in a wider period-eccentricity distribution, and, in particular, could explain the formation of long-period binaries with moderate or low eccentricities
\citep{2024ApJ...972L..18H, 2025arXiv250705602B}. 
Furthermore, as mentioned before, the electromagnetic rocket mechanism could naturally explain the observed trend towards spin-kick alignment.

Finally, we have to mention the early suggestion that the neutrino emission asymmetry can explain the high velocities of NSs. This mechanism was proposed in a brief note by
\citet{1984SvAL...10...87C}. 

As most of the gravitational energy liberated during a core-collapse SN, $E_\nu\sim 10^{53}$~erg  is carried away by neutrinos, even a small asymmetry $\sim 1$~\% can result in a huge impact. Since the neutrino momentum is $p = E/c$, the natal kick from anisotropic emission is
\begin{equation}
v_\mathrm{NS} = \alpha (E_\nu/c) / M_\mathrm{NS} = 120 \, \mathrm{km\, s}^{-1}\ (\alpha/0.01) (E/10^{53} \mathrm{erg}) (1.4 \mathrm{M}_\odot / M_\mathrm{NS}),
\end{equation}
where $\alpha$ is the degree of emission asymmetry.

Such an emission asymmetry can come about from a variety of mechanisms. During the first few hundred milliseconds of an explosion, it can arise due to asymmetric accretion onto the proto-neutron star, which results in neutrino emission hotspots \citep{2013A&A...552A.126W}. The direction of the anisotropy is expected to vary with time, with no or little preference for alignment with the rotation axis.   Another
mechanism for an emission asymmetry between
electron neutrinos and electron antineutrinos is the
Lepton-number Emission Self-sustained Asymmetry (LESA) instability in the proto-neutron star, a low-mode flow instability that produces a dipolar lepton number asymmetry in the proto-neutron star convection zone \citep{tamborra_14a}. Sufficiently long self-consistent simulations to evaluate these neutrino-driven kicks have only become available more recently, see Section~\ref{hydro}.

Large emission asymmetries can also occur if the proto-neutron star has a strong and highly asymmetric magnetic field. 
The idea of this mechanism is based on the polarization of electrons (and positrons) in the magnetic field of a proto-neutron star \citep{lai_98}. 
Due to parity violation in weak interactions, neutrino production processes involving particles aligned with the magnetic field are emitted with a preferential direction.
If all electrons and positrons are polarized, then the expected velocity depends on the magnetic field as $v\sim 30 B_{14}$~km~s$^{-1}$, where $B_{14}$ is the magnetic field in units of $10^{14}$ G. Thus, for a reasonable value of the magnetic field, it is possible to obtain a velocity of a few hundred km~s$^{-1}$.  Unlike anisotropic neutrino emission due to accretion hotspots, the magnetically-induced emission anisotropy could provide sustained acceleration in a fixed direction aligned with the rotation axis and act on longer time scales, similar in some respects to other rocket-type mechanisms.

\subsubsection{Modern simulations}\label{hydro}
This subsection briefly describes several representative examples of supernova explosion calculations performed in recent years by different groups. We do not intend to provide a general description of the various models. Such reviews can be found e.g., in \cite{2007PhR...442...38J, 2018ASSL..457....1C, 2017hsn..book.1575J,2020LRCA....6....3M}. Instead, we focus on calculations of the kick velocity.
We start with scenarios focused on hydrodynamics and a detailed treatment of neutrino transport, including outer layers of ejecta and fallback.

The necessity of analyzing various non-radial asymmetries in SN explosions motivated the study of NS kicks by
\cite{2006A&A...457..963S}. 
In their axisymmetric (2D) models of neutrino-driven explosions, hydrodynamic instabilities are responsible for the asymmetry and kick velocity. 
Very rapid rotation, as well as the magnetic field, jets, and exotic neutrino processes, are not considered. Thus, the model presents a rather conservative approach that is still sufficient to explain important properties of explosions and remnants. 

The authors observed that in many explosions, a significant dipole ($l=1$) asymmetry can develop due to the two key hydrodynamic instabilities that operate before and around the onset of the explosion, i.e., neutrino-driven convection \citep{herant_94} and the standing accretion shock instability \citep{blondin_03}.
The interaction between the NS and the slower (and denser) part of the ejecta is mediated by gravitational force and is sometimes called the
``gravitational tugboat mechanism''  \citep{2013A&A...552A.126W}, although some authors prefer
a more nuanced terminology with less emphasis on the role of gravity
in transferring momentum to the NS \citep{2024ApJ...963...63B}.  
As the denser parts of the ejecta exert a net pull on the NS and accelerate
it in their direction, they experience acceleration in the opposite direction, resulting in an opposite net momentum of the NS and the overall ejecta in compliance with momentum conservation. 
The slower (and denser) portions of the ejecta may even fall back onto the NS, to further accelerate the compact object in their direction and away from the rapidly ejected unbound material \citep{2013A&A...552A.126W} as expected from momentum conservation.
(Of course, the internal gravitational interaction of material that ultimately ends up inside the NS cannot produce a net acceleration, which is why some authors eschew the term ``tugboat'' and focus on the ejecta momenta as these determine the NS momentum.)
The initial large-scale density asymmetry that accelerates the NS by gravitational forces comes about due to the buoyancy-driven instability.
In some sense, this mechanism can be viewed as the late, highly non-linear phase of a buoyancy-driven instability that magnifies density and velocity anisotropies between rising low-density bubbles and the denser surrounding material. 

In the code by \citet{2006A&A...457..963S}, an NS formally has infinite inertia. Thus, it does not move, i.e., it always stays at the origin of coordinates. However, careful calculations of the momentum of the ejected material allow for the derivation of the kick velocity, and
\citet{2006A&A...457..963S} also performed simulations in an accelerated reference frame to ensure the results are robust when the acceleration of the NS is taken into account. The extrapolation of NSs velocities beyond the time of calculation, about one second, suggested a bimodal velocity distribution in correspondence with the one derived by \cite{2002ApJ...568..289A} from observations. The higher velocities are due to the development of the dipole mode instability during the explosion. 
Simulations with Cartesian codes without restrictions on the motion of the NS confirmed the physics of the hydrodynamics kick mechanism
\citep{nordhaus_12}.

Another 2D study that specifically focused on kick velocity calculations was presented by
\cite{2019PASJ...71...98N}
using isotropic diffusion source approximations \citep{liebendoerfer_09}. 
These authors presented 2D hydrodynamical calculations for a large set of progenitors with different masses and so with different compactness and energy of explosion. In many respects, their approach is similar to that used by the Garching group.
In particular, the kick velocity is estimated based on the asymmetry of the momentum distribution of the ejected gas relative to $z$-axis (which is the axis of symmetry):
\begin{equation}
    V_\mathrm{kick}=\alpha_\mathrm{gas} P_\mathrm{gas}/M_\mathrm{NS}.
\end{equation}
Here $\alpha_\mathrm{gas}=|P_\mathrm{z, gas}|/P_\mathrm{gas}=|\int{\mathrm{d}m\, v_\mathrm{z}}|/\int{\mathrm{d}m\, |\vec{v}|}$ and $P_\mathrm{z, gas}$ is the $z$-component of the total momentum $P_\mathrm{gas}$.

The gas momentum depends on the total energy of the ejecta (kinetic, internal and gravitational), which by itself is related to progenitor compactness $\xi$:
\begin{equation}
    \xi=\frac{M/\mathrm{M}_\odot}{R(M)/1000 \, \mathrm{km}}.
\end{equation}
The compactness grows with increasing progenitor mass up to $17\, \mathrm{M}_\odot$ and then decreases. Larger compactness results in greater energy deposited in the ejecta. The asymmetry parameter $\alpha_\mathrm{gas}$ at $t>2$~s mostly varies in the narrow range $\lesssim 0.3$.  Thus, the largest energy and kick velocity at the end of simulations, $\sim 850$~km~s$^{-1}$, is reached for progenitors with $M=17 \, \mathrm{M}_\odot$.  More typical velocities are $\sim 100$~--~$400$~km~s$^{-1}$. The final velocity can be larger as the velocity continues growing when the calculations are terminated. 
Interestingly, the authors obtained a correlation: more massive NSs have larger velocities. 

For understanding the typical scale of hydrodynamic kicks, it should be noted that  around the time when
the kicks reach their final values and accretion ceases, the explosion energy $ E_\mathrm{expl}$ is still mostly contained in the thermal energy of the neutrino-heated bubbles.  Hence, the
momentum of the ejecta is considerably smaller than the characteristic
momentum scale corresponding to the explosion energy and the mass of the shocked ejecta $M_\mathrm{sh}$ around the time when accretion stops.
$M_\mathrm{sh}$ is typically only of order $1\mathrm{M}_\odot$, again much smaller than the final ejecta mass. Thus, in terms of
these characteristic parameters \citep{VignaGomez:2018},
\begin{equation}
    V_\mathrm{kick}\sim 0.08 \sqrt{2 E_\mathrm{expl} M_\mathrm{sh}}/M_\mathrm{NS}.
\end{equation}

Kick velocity calculations based on self-consistent multi-group neutrino transport simulations were long restricted to axisymmetry. 
This is so because it is necessary to run the simulation until several seconds after the bounce. Such modelling for many progenitors in the 3D case is computationally very expensive. However, the first 3D simulations with
\emph{grey} neutrino transport  were carried out
over a decade ago
\citep{wongwathanarat_10b,2013A&A...552A.126W}. These models confirmed that high kick velocities of several $100\,\mathrm{km}\,\mathrm{s}^{-1}$ remain possible in 3D and were not mere artefacts of the artificial assumption of axisymmetry. Kicks were first studied using self-consistent, relativistic 3D simulations
with multi-group transport by
\citet{2017MNRAS.472..491M,mueller_19a}. Compared to the parameterised 3D models of 
\citet{2013A&A...552A.126W}, \citet{mueller_19a} obtained even higher extrapolated final kicks of more than $1000\,\mathrm{km}\,\mathrm{s}^{-1}$. Moreover, their 3D models indicated -- in fact, even before the large set of 2D simulations by \citet{2019PASJ...71...98N} -- a loose correlation between the kick and the explosion energy and progenitor core mass.

Recently, the Garching group also presented 3D calculations with multi-group transport with the \verb|Prometheus-Vertex/NEMESIS| code, and 
with more and more advanced analyses related to the origin of the kick velocity of NSs \citep{JankaKresse2024}. 
They present 3D calculations for eight stellar masses: from 9 up to $75\, \mathrm{M}_\odot$. Six progenitors produce NSs, while two produce BHs. For some masses, several variants of calculations are performed. The simulations demonstrate the evolution of the compact object and ejecta up to $\sim 8$~s after the core bounce. This time is sufficient for the appearance and development of hydrodynamical instabilities in the post-shock material. Thus, the kick velocity can be readily calculated.  

Typically, the acceleration of the NS takes place during a few seconds (mostly after $\sim 0.5$~s) when the gas accretion onto the proto-NS is not important. Late fallback on the scale of minutes to hours (and longer) has very little effect on the kick. 

To calculate the kick due to neutrinos, the effects of neutrino emission, absorption, and scattering in material outside the neutrinosphere are taken into account. Still, as in earlier modeling, the hydrodynamical kick dominates for all massive stars. Only for their models with progenitor masses $9\, \mathrm{M}_\odot$ and $9.6\, \mathrm{M}_\odot$  do neutrinos dominate, but the total velocity in these cases is $\lesssim 60$~km~s$^{-1}$. The largest NS kick obtained for the progenitor mass $ 20 \, \mathrm{M}_\odot$ is $\gtrsim 1300$~km~s$^{-1}$.

Recently, 
\cite{2024ApJ...963...63B} 
presented twenty 3D simulations of core-collapse SNe for different progenitor masses.\footnote{A detailed analysis of correlations found in these calculations is presented in \cite{2024ApJ...964L..16B}.} 
The authors used the code \verb|FORNAX|.
 They combine progenitor models from
two recent sets of solar-metallicity stellar evolution calculations \citep{2016ApJ...821...38S,2018ApJ...860...93S},
which differ slightly in terms of stellar microphysics and initial solar composition, which is taken from \citet[][$Z=0.0133$]{2003ApJ...591.1220L} and \citet[][$Z=0.0134$]{2009ARA&A..47..481A}, respectively.

 General relativity effects are taken into account approximately. Their model does not yet account for the rapid rotation of progenitors and magnetic field effects. 
Their study focuses on kick and spin properties. One of the main conclusions reached by the authors coincides with previous studies presented above. Progenitors with smaller compactness produce low-velocity compact objects, and in cases of higher compactness, the final velocity is higher. This feature again leads to a positive correlation between the NS mass and kick velocity: less massive NSs have smaller velocities of $\sim 100$~--~200~km~s$^{-1}$. 
Neutrinos provide only a modest contribution $\lesssim 100$~km~s$^{-1}$. 

None of the aforementioned simulations has yet investigated the potential impact of magnetic fields on explosion asymmetries and kicks. However,
first tentative magnetohydrodynamic simulations of neutrino-driven supernovae \citep{2025PhRvD.111f3042S} only suggest a minor impact of magnetic fields in the case of non-rotating progenitor stars. Kicks in
magnetorotational explosions were recently investigated by \citet{2023MNRAS.522.6070P} using
3D simulations. Although such explosions may be both strongly asymmetric and substantially more energetic than neutrino-driven ones, \citet{2023MNRAS.522.6070P} only find kicks of up to $250\,\mathrm{km}\,\mathrm{s}^{-1}$. These relatively modest kicks occur because the simulated magnetorotational explosions exhibit bipolar rather than unipolar asymmetry so that
the momentum of the ejecta almost cancels.
Interestingly, the kicks are not aligned with the rotation axis because the bipolar outflows do not remain perfectly aligned with the rotation axis due to instabilities; hence, the ejecta develop some net momentum perpendicular to the rotation axis. Three-dimensional simulations remain too expensive to extensively explore the parameter space with the long simulation times required to confidently predict kick velocities.
Recent 2D simulations by \citet{2024PhRvD.110h3025K} have indicated a strong hemispheric asymmetry between the
jets in some instances, which permits kicks of several $100\,\mathrm{km}\,\mathrm{s}^{-1}$. The phenomenology of kicks in magnetohydrodynamically-driven explosions still needs to be explored further. Because magnetorotational explosions are expected to be rare, their role in shaping the kick distribution of compact objects is likely minor, however.



The theory of hydrodynamic kicks predicts that special evolutionary channels of NS formation should lead to kick distributions distinctly different from those of normal core-collapse supernovae. 
Particularly interesting formation  scenarios include  the
electron-capture supernova (ECSN)
channel, which was first investigated in depth in the 1980s
\citep{nomoto_84,nomoto_87}, and the related scenario of accretion-induced collapse (AIC) as laid
out in the seminal works of \citet{1990ApJ...353..159B} and \citet{1991ApJ...367L..19N}.
ECSNe are speculated to occur in stars
just below the lower mass limit for normal core-collapse supernovae when
electron captures on neon and magnesium trigger collapse on a dynamical time scale after carbon burning. In this scenario the collapsing core is of low mass and surrounded by a dilute envelope, limiting the amount of mass that can be accreted or ejected asymmetrically. In AIC, collapse occurs due to electron captures in an accreting O-Ne white dwarf, again implying that only a small amount of mass can be ejected asymmetrically.
NSs born in 
ECSNe or by AIC are therefore predicted to experience small hydrodynamic kicks \citep{2013A&A...558A..39T,Gessner:2018}. In these scenarios, accretion onto the NS terminates very quickly as the explosion develops because the density in the shells surrounding the core drops steeply with radius (in ECSNe) or because there is simply little mass outside the NS that can be accreted (in AICs). Furthermore, the rapid onset of the explosion in ECSNe and AICs leaves less time for large-scale hydrodynamic instabilities to grow and hence results in less asymmetric ejecta. For this reason, hydrodynamic kicks in simulations of ECSNe
are only of the order of a few $\mathrm{km}\,\mathrm{s}^{-1}$ \citep{Gessner:2018}.
However, such NSs may still receive substantial kicks due to other mechanisms. In particular, even without strong magnetic fields, neutrino-induced kicks of several $10\,\mathrm{km}\,\mathrm{s}^{-1}$ can still be attained \citep{JankaKresse2024} due to the LESA.


\subsection{Black holes}
It has been realised for over a decade that some of the kick mechanisms for NSs can also impart sizeable kicks onto BHs. As first pointed out by \citet{janka_13}, hydrodynamic kicks of similar velocity as for NSs can be attained for BHs if the BH is formed by \emph{fallback} in an explosion with partial mass ejection.
A BH can be formed by fallback either because accretion onto the NS continues after the explosion develops, or because initially outward-moving material is eventually decelerated and cannot escape the potential of the compact object.
As substantially more mass can be accreted asymmetrically on a BH than on an NS in such an explosion, the momentum imparted on the BH can be significantly bigger, compensating for the higher remnant mass and perhaps lower initial ejecta velocities in such BH-forming explosions.  
The  analytic argument of \citet{janka_13} was subsequently largely borne out by multi-dimensional simulations. 
\citet{chan_18,2020MNRAS.495.3751C} 
simulated BH formation in 
$40\,\mathrm{M}_\odot$ stars by fallback in 3D
and followed the subsequent evolution
of the explosion 
beyond shock breakout from the stellar surface.
They demonstrated that fallback explosions can produce BHs moving at several $100\, \mathrm{km}\,\mathrm{s}^{-1}$ provided that a large fraction of the
stellar envelope can still be ejected. In the case of a less energetic explosion with fallback of the entire helium core and
part of the hydrogen envelope, \citet{chan_18} found that a sizable BH kick builds up initially, but is eventually neutralized by further fallback of envelope material, which results in a rather spherical explosion.

Subsequent simulations in 2D \citep{2022MNRAS.512.4503R} and 3D \citep{JankaKresse2024,2024ApJ...963...63B} further point to a dichotomy in outcomes as seen in the models of \citet{chan_18,2020MNRAS.495.3751C}. Depending on whether or not a sufficiently energetic explosion is already underway by BH formation via fallback, the hydrodynamic kick on the BH may either be large or negligible. The timing of shock revival and the energy of BH-forming explosions are likely sensitive to details of the neutrino transport and the nuclear equation of state, so that first-principle predictions for the prevalence of these two channels are not yet forthcoming. A significant challenge in modelling the outcomes of BH formation in the fallback scenario consists in the long time scales.
Hydrodynamic BH kicks may only asymptote to their final values
several minutes after the explosion \citep{2020MNRAS.495.3751C}, significantly after BH formation and beyond the usual simulation time of most core-collapse supernova explosion models.
The situation is even more extreme for BH spins,
since even small amounts of late-time fallback can substantially
alter the BH spin.
However, the physics exemplified by the simulations suggests that
large kicks may be easiest to obtain if the fallback is substantial but far from complete, producing BHs of a few solar masses in
what was previously \citep{2012ApJ...757...91B} thought to be a BH ``mass gap''.
Phenomenological models for BH kicks based on the physics of fallback seen in multi-dimensional simulations \citep{MandelMueller:2020} indeed exhibit such a pattern.

Although unipolar asymmetries conducive to substantial BH kicks emerge naturally in neutrino-driven fallback explosions, the attainable BH kicks in bipolar explosions in the collapsar scenario are much less certain.
In this scenario, there is no explosion prior to BH formation, but asymmetric mass ejection can occur later when an accretion disk is formed and outflows from the jet are launched via jets or disk winds \citep{macfadyen_99}.
The recent work of
\citet{Halevi:2025} followed the collapse of a rapidly rotating collapsar progenitor in 3D and obtained a kick of $72\,\mathrm{km}\,\mathrm{s}^{-1}$ immediately after BH formation, but without following the accretion onto the BH 
and the mass ejection from the collapsar disk, the kicks in the collapsar scenario remain very much in uncharted territory.

Regardless of whether BH formation is associated with
an explosion or not, the BH will acquire a kick from
anisotropic neutrino emission during the proto-neutron star
phase that precedes the collapse. However, the kick due to neutrinos
is expected to be small. The amount of energy radiated as neutrinos
is limited to a fraction of the binding energy corresponding to the maximum NS  mass, and the time-averaged emission anisotropy is likely lower in cases of quiet BH formation  without shock revival after a transient proto-neutron star phase, as stable, long-lasting unipolar accretion asymmetries only emerge during
the explosion phases. Consequently, 
\citet{JankaKresse2024} and \citet{2024ApJ...963...63B} find small
neutrino-induced kicks of $10\,\mathrm{km}\,\mathrm{s}^{-1}$ or at most a few $10\,\mathrm{km}\,\mathrm{s}^{-1}$ for BHs shortly
after formation, which then decrease with time as the BH mass grows. Neutrino emission after BH formation drops precipitously \citep{2025MNRAS.538..572S} so that emission asymmetries during this phase cannot substantially add to the kick anymore.

\section{Direct observational constraints on kicks }
\label{s:direct_constraint}

\subsection{Single radio pulsars}\label{s:pulsars}


Radio pulsars were discovered as sources of periodic radio signals by \cite{HewishBell1968Natur}. For most radio pulsars, the signal is repeated with a precise period ranging from a few milliseconds to tens of seconds. Radio pulsars are uniquely attributed to NSs following the lighthouse model \citep{Lorimer2004hpa}. Within this model, the period of radio signal repetition is associated with the spin period of the NS. The beamed radio emission is produced in the magnetosphere and is only accessible to observations when the emission beam points towards the Earth.   


Soon after the discovery of isolated radio pulsars, it was noticed that these objects have large peculiar velocities, and their spatial distribution 
has a large scale height above the Galactic plane in comparison to their progenitors -- massive stars  \citep{GunnOstriker1970ApJ}. As we explained in the introduction, these are two indications of large spatial velocities received during or after the supernova explosion.

It is worth discussing what is actually measured for isolated radio pulsars and how these measurements are related to their spatial velocity. Unlike normal stars, radio pulsars have no detected absorption or emission lines produced in their atmospheres\footnote{Some young thermally emitting NSs have absorption features possibly associated with their uncertain magnetic fields. These measurements cannot be used to constrain the radial velocity.}. Pulsar radio emission is produced via a coherent non-thermal mechanism far above the surface by electrons and positrons moving near the speed of light \citep[see, e.g.,][]{Chernoglazov2024ApJ}. Absorption lines present in their radio spectra are usually interpreted as redshifted absorption lines caused by the propagation of the radio signal through clouds of Galactic neutral hydrogen (HI). These features are occasionally used to constrain distance to radio pulsars, see e.g.~the discussion by \cite{FrailWeisberg1990AJ}. Thus, there is fundamentally no directly measured constraint on the radial velocity of isolated radio pulsars. The only case in which measured constraints can be obtained is that of a binary with a radio pulsar. In this case, the optical spectrum of a normal star can be used to constrain the center of mass radial velocity.

Geometrically, the radial velocity might not be very special. It is frequently assumed that the peculiar velocity is drawn from an isotropic distribution, so the radial component for a sample of radio pulsars is drawn from exactly the same distribution as the other two components. This assumption was recently challenged by \cite{2023ApJ...944..153M} who discussed the role of alignment between the rotational axis and the orientation of the kick, see Section~\ref{s:kick_spin} for more details. 

As is common in astronomy, the transverse velocity measurement for radio pulsars is actually a combination of proper motion and distance measurements. Proper motions of pulsars have been measured with good precision for almost half a century \citep{1977ApJ...213L...1H}. Currently, such measurements are available for hundreds of pulsars; see the ATNF catalogue\footnote{https://www.atnf.csiro.au/people/pulsar/psrcat/} \citep{2005AJ....129.1993M}. 
Direct distance measurements via parallax are particularly challenging for isolated radio pulsars. 
Individual radio telescopes -- even as large as the Effelsberg -- have angular resolution worse than $\approx 10'$ at a frequency $\sim 1$ GHz. 
Even the closest pulsars are located at large distances $\sim 100$~pc~--~$1000$~pc, which requires angular resolution better than 1~mas for precise parallax measurements. 
Thus, the most precise parallax measurements are made using very long baseline interferometry. The first reliable measurements were performed by \citet{Chatterjee2001ApJ}.  
Presently, many new measurements are made with VLBA \citep{Deller:2019} and MeerKat \citep{2024MNRAS.530..287S}.  The Square Kilometer Array will provide parallax measurements 
for a large sample of isolated radio pulsars \citep{2011A&A...528A.108S}.  In the meantime, distances are frequently estimated based on the dispersion measure of the pulse, despite the significant scatter and possible bias in such estimates \citep{Deller:2019}.

Obviously, the NS natal velocity distribution is not equivalent to the present-day distribution of the spatial velocities, especially for a selected group of sources, e.g., isolated radio pulsars with measured proper motions and parallaxes. In a simplified approach, it is possible to use just very young pulsars for which the effects of movement through the Galactic gravitational potential can be neglected. Then, even without corrections for Galactic rotation (if the pulsar is relatively nearby), the combination of proper motion and parallax provides the transverse velocity, which is a good proxy for the spatial velocity.  
Many proposed kick distributions are based on this approach. However, such samples are inevitably small and biased. 
 Thus, the kick distribution is often derived in a model-dependent way using a larger set of observed sources. 
 Typically, it is done via population synthesis modelling, see, e.g., \cite{2007PhyU...50.1123P}. 
 In this approach, initial parameters of NSs are specified (spin period, magnetic field, birthplaces, velocity, etc.). Then, the evolution of NSs is modelled according to prescribed laws and parameters (spin-down rate, magnetic field evolution, gravitational potential of the Galaxy, etc.). To account for various selection effects, it is necessary to model the detectability of pulsars in specific surveys. This requires assumptions about the radio luminosity, dispersion measure, beaming, survey sensitivity, and sky coverage.
 Finally, properties of ``artificially detected'' pulsars are confronted against the observed ones. 
 When the two samples match, one can conclude that the initial parameters in the simulations, particularly the kick distribution, match their values in nature.  Below, we describe the results based on both methods. 

In this paper, we do not intend to provide a comprehensive historical review of numerous studies that have aimed to derive the kick velocity distribution of NSs based on radio pulsar data. Still, in the following, we describe and discuss the results of several illustrative studies in historical order.

A major breakthrough in studying the kick velocities of radio pulsars was made by \cite{1994Natur.369..127L}. 
 With a new model of the Galactic electron density distribution (\citealt{1993ApJ...411..674T}), these authors
demonstrated that, on average, distances to radio pulsars have been significantly underestimated previously.
As a result, the spatial velocities of the pulsars increased in comparison with previous estimates by a factor of $\sim 2$--3. Using a sample of 29 sources with measured proper motions and ages $<3$~Myr they obtained a mean 3D velocity of $\sim 450\pm90$~km~s$^ {-1}$. Since that study, the typical kick velocities of most radio pulsars are known to be several hundred~km~s$^ {-1}$. 

 Since the late 1990s, several groups based on different approaches have proposed bimodal distributions of the kick velocity, including \citet{1998ApJ...505..315C} and \citet{1998ApJ...496..333F}. 
 A detailed population synthesis analysis of radio pulsars was presented by
\cite{2002ApJ...568..289A}. 
Their bimodal distribution consisted of two gaussian components with $\sigma_\mathrm{v1}=90$~km~s$^ {-1}$ and  $\sigma_\mathrm{v2}=450$~km~s$^ {-1}$. About 40\% of pulsars belong to the low-velocity part of the distribution. 
Note that this bimodal distribution was obtained just by fitting the data on isolated radio pulsars. 
Thus, the low-velocity component cannot be explained by electron-capture SNe 
\citep{2002ApJ...574..364P}. 

For approximately 15 years, the distribution of \cite{2002ApJ...568..289A} was one of the most popular choices of the kick velocity. Another, even more popular distribution was proposed by
\cite{2005MNRAS.360..974H}. 
This is a single-mode Maxwellian distribution for the 3D speed $ f(v)\propto v^2\exp{\left(-\frac{v^2}{2\sigma^2}\right)}$ with $\sigma\approx 265$~km~s$^ {-1}$ (mean velocity $\approx 423$~km~s$^ {-1}$).
The authors did not conduct population synthesis modelling to follow evolution in the Galactic potential. Instead, they analyzed properties of 233 pulsars with measured proper motions. 
After selecting objects with characteristic ages $<3$~Myr, the sample contains 73 pulsars with one-dimensional and 46 pulsars with two-dimensional velocity measurements. The proposed Maxwellian distribution was found to fit the underlying 3D velocity distribution for both subsamples. 

The distribution by \cite{2005MNRAS.360..974H} has very few low-velocity objects, which is in tension with many other estimates. Recently, the issue was clarified by
\cite{2025arXiv250522102D}. 
These authors demonstrate that the analysis by \cite{2005MNRAS.360..974H} contains an error due to missing a Jacobian needed to correct for the logarithmic histogram bin sizes. 

A very detailed population synthesis of radio pulsars was carried out by
\cite{2006ApJ...643..332F}. 
The results of modelling were compared with properties of a large sample of pulsars detected in Parkes and Swinburne Multibeam surveys.
In particular, the authors studied several variants of the kick velocity distribution. 
They advocate a single-mode exponential distribution. The probability for each velocity component $v_i$ can be written as
\begin{equation}
    p(v_i) = \frac1{2 \langle v_i \rangle} \exp{\left( - \frac{|v_i|}{\langle v_i \rangle}\right)}.
\end{equation}
The mean 3D velocity is $380^{+40}_{-60}$~km~s$^ {-1}$.
This velocity distribution was selected on the basis of proper motion measurements and then used in the population synthesis calculations. The authors checked how alternative models of the kick velocity distribution affect their population modelling results and found that models that fit the proper motion data similarly well also produce very similar predictions for other population properties. 

A bimodal velocity distribution with two Maxwellian components was advocated in a series of papers by Andrei Igoshev and co-authors:
\cite{2017A&A...608A..57V}, 
\cite{2020MNRAS.494.3663I}, 
\cite{2021MNRAS.508.3345I}. 

\cite{2017A&A...608A..57V} analyzed 28 pulsars (19 among them with ages $<10$~Myr) with measured parallaxes and proper motions. Their results strongly favor a bimodal velocity distribution described by two Maxwellians with the parameters very similar to those of \cite{2002ApJ...568..289A}: $\sigma_1=60$-95~km~s$^ {-1}$, $\sigma_2=276$-368~km~s$^ {-1}$, and the fraction of the low-velocity pulsars $w=0.3$-0.52. 

This result was refined by \cite{2020MNRAS.494.3663I}, who used a larger data set of pulsars with well-measured proper motion and parallax. With a sample of 69 objects (21 of them with characteristic age $<3$~Myr) it was shown that the bimodal distribution fits the properties of radio pulsars. The distribution is defined by the following parameters: $\sigma_1=100$-150~km~s$^ {-1}$, $\sigma_2=270$-326~km~s$^ {-1}$, and  $w=0.27$-0.59. These values correspond to the whole sample of the analyzed pulsars, including older ones. Thus, the influence of evolution in the Galactic potential has not been taken into account. If using just the youngest pulsars with characteristic ages $<3$~Myr, for which the effects of the Galactic potential are negligible, the values are: $\sigma_1=41$-81~km~s$^ {-1}$, $\sigma_2=311$-381~km~s$^ {-1}$, and  $w=0.1$-0.31.
Among young pulsars, the fraction of objects with velocity $<60$~km~s$^ {-1}$ is about 2--8\%.

The estimates above are based on the velocity measurements for isolated radio pulsars. 
However, the majority of their progenitors (massive stars) are born in binary (or multiple) systems \citep{MoeStefano2017ApJS}. 
If the natal kick is low, then with a high probability the binary remains bound. Thus, the population of isolated pulsars can have a deficit of low-velocity objects. To clarify this issue, it is necessary to derive the kick velocity distribution on the basis of a joint analysis of isolated radio pulsars and binary systems with massive stars. This has been done by \cite{2021MNRAS.508.3345I}.
These authors generally confirm the results by \cite{2020MNRAS.494.3663I}. 

\cite{2021MNRAS.508.3345I} performed population synthesis of Be X-ray binaries varying the kick velocity distribution in the range defined by the radio pulsar properties. This allows probing the low-velocity part of the distribution with better precision. The authors obtained 
$\sigma_1=30$-80~km~s$^{-1}$ and  $w=0.1$-0.3, while the parameter of the high-velocity part of the distribution was assumed to be the same as in \cite{2020MNRAS.494.3663I}: $\sigma_2=336$~km~s$^{-1}$.
In the optimal model, the authors do not use a specific velocity for electron-capture SN. Thus, the contribution from this channel of NS formation is ``hidden'' in the low-velocity part of the distribution. 
As electron-capture SNe might be more numerous in interacting binary systems and various types of interaction between the companions can modify the natal kick velocity, formally, we cannot expect that the same kick distribution can be applied to all NSs born from different progenitors. Still, the results by \cite{2021MNRAS.508.3345I} suggest that in many cases the difference is within the range of uncertainties.

A recent analysis of the properties of Be X-ray binaries suggests a component of the kick distribution with a very low velocity $\lesssim 10$~km~s$^{-1}$ \citep{Valli:2025}. In addition, the authors claim that a narrow component with a typical velocity $\sim100$~km~s$^{-1}$ (which approximately corresponds to the low-velocity component from \citealt{2021MNRAS.508.3345I}) is nearly aligned (within $5^\circ$) with the spin axis of the progenitor.



\cite{2024A&A...687A.272D} 
and
\cite{2024A&A...689A.348D} 
analyzed older pulsars while accounting for the dynamical effect of the Galactic potential, allowing them to infer their natal kick velocities.
The authors accurately calculate the trajectories of NSs in the Galactic gravitational potential after they received a kick (see details in \citealt{2024A&A...689A.348D}).  Inference on the kick distribution is based on the parameters of the Galactic orbits of NSs, especially on their eccentricities. As the orbits  in the Galactic potential are not Keplerian, the authors define the eccentricity as
\begin{equation}
    \tilde e=\frac{R_\mathrm{max}-R_\mathrm{min}}{R_\mathrm{max}+R_\mathrm{min}},
\end{equation}
where $R_\mathrm{max}$ and $R_\mathrm{min}$ are the maximum and minimum galactocentric distances, respectively. Zero kicks leave $\tilde e=0$ while large kicks cause pulsars to escape the Galaxy,  $\tilde e \rightarrow 1$. A large eccentricity can also be obtained if the kick nearly compensates for the initial circular velocity. Finally, in very rare cases, an NS can have zero eccentricity after receiving a kick equal to twice the circular velocity directed against the initial motion.

For each pulsar with measured parallax and proper motion, the authors calculate the trajectory and the corresponding probability distribution for the eccentricity. Then, these results are confronted with calculations of eccentricities (as a function of age) for a given kick. Finally, for each observed pulsar, a probability distribution for its natal kick is obtained.  This makes it possible to use data on pulsars older than 10 Myr; pulsars younger than 10 Myr are not significantly affected by dynamics according to \cite{2024A&A...689A.348D}.  


\cite{2025arXiv250522102D} find that there is no conclusive evidence for a bimodal distribution in either the young pulsar sample, once it is extended beyond the systems used by \cite{2020MNRAS.494.3663I}, or in the sample of older pulsars \citep{2025arXiv250301429D}.  They thus conclude that the two modes in the apparent velocities of young pulsars found by \cite{2020MNRAS.494.3663I} could be a consequence of low-number statistics.  They fit the probability distribution of the initial velocities of isolated pulsars with a single log-normal velocity distribution of the form \citep{2025arXiv250301429D}
\begin{equation}
    p(v|\mu, \sigma)=\frac1{v\sigma \sqrt{2\pi}}\exp\left({-\frac{(\ln v -\mu)^2}{2\sigma^2}}\right).
\end{equation}
The best fit for pulsars younger than 10 Myr is obtained for $\mu=5.60$ and $\sigma=0.68$. 
\cite{2025arXiv250522102D} demonstrated that the results of \cite{2005MNRAS.360..974H} are inconsistent with the pulsar data because of the missing Jacobian in the latter work; once that Jacobian is correctly included, the single Maxwellian, double Maxwellian, and log-normal distributions are all consistent with observations.

The \cite{2025arXiv250522102D} distribution of single pulsar velocities probably underpredicts the number of low-velocity NSs in view of the abundance of NSs in wide binaries and in globular clusters with escape velocities under $50$~km~s$^{-1}$, since fewer than $1$\% of the inferred distribution falls in the range  $<50$~km~s$^{-1}$.  However, truly single massive stars are very rare: the vast majority are found in binaries or higher order multiples \citep{MoeStefano2017ApJS}, with many interacting before the first supernova \citep{Sana2012Sci}.  This imposes selection effects that determine which compact remnants are single: they are ones that received a large enough kick velocity to be ejected from binaries.  This could partly explain why slowly kicked NSs are rare in the single-pulsar population.

Moreover, binary interactions change progenitor properties, including mass and structure, and thus both explodability \citep{Maltsev:2025} and perhaps natal kicks  (see also Section~2.1 about electron-capture SNe). Thus, a possible explanation for the discrepancy between low kicks inferred from observations of isolated pulsars, on the one hand, and the need for a population of low kicks from NS binaries as well as NS prevalence in clusters, on the other hand, is that there is a subpopulation of NSs that are formed specifically as a consequence of binary interaction. They receive low kicks and stay in binaries, thus never appearing in the single-pulsar population.  For example, \cite{Willcox:2021} conjectured that low natal kicks might be imparted in electron-capture SNae that exclusively or predominantly occur for stars stripped by mass transfer, thus avoiding a second dredge-up.  Other binary interactions, such as accretion-induced collapse, could be responsible for a special subpopulation of NSs that receive very low natal kicks and overwhelmingly remain in binaries.  An alternative possibility is that NSs that get low kicks are somehow different from normal pulsars and avoid detection in pulsar surveys unless they are recycled by accretion.


%


\subsection{Location within SN remnants}

 In this subsection, we discuss the velocities of NSs observed in SN remnants (SNRs). 
Sources can belong to different classes: radio pulsars (PSRs), magnetars (anomalous X-ray pulsars -- AXPs, and soft gamma-ray repeaters -- SGRs), and central compact objects in SNRs (CCOs), see Table~\ref{tab:SNR}. 
In the context of kick determination, the location of a compact object within an SNR (or in a pulsar wind nebula -- PWN) provides several advantages compared to other NSs. Firstly, the object is definitely young, so its velocity could not have changed since birth. If it can be established (by the absence of the secondary star) that the NS was not born in a binary system, complications related to the evolution in a binary can be omitted. Though, 
the current supernova could be the second supernova in a binary that was disrupted by the first supernova, e.g., \cite{Hirai:2020}. If an NS is related to an SNR, then it is possible to obtain an independent estimate of the age. In the case of SNRs, it is possible to obtain an independent (and often precise!) measurement of the distance. Analysis of the morphology of the remnant can provide additional information for the reconstruction of the radial component of the velocity.  Finally, measurements of the kinematics of the ejecta give information about the SN explosion. 

 In the case of SNRs, several estimates of NS velocities were obtained without proper motion measurements. This is possible if one can locate the location of the explosion. In the simplest case of a circular remnant, it is possible to assume that the explosion occurred in the geometrical center. Then, with known age and distance to the remnant, simply from the measurement of the shift between the NS position and the point of explosion, one obtains the 2D velocity.  However, many remnants do not have a circular shape. Also, the place of the explosion can be offset from the geometric center of the remnant because the SN progenitor produced an asymmetric cavity due to its wind while moving in the ISM \citep{2000astro.ph..5572G}. 
 Thus, velocity estimates obtained without proper motion measurements are not very robust. 
In addition, the absence of proper motion measurements increases the probability of an erroneous association of the NS and the SNR, as in the case of SGR J1900-14 \citep{1994ApJ...431L..35V} which resulted in an unrealistic velocity estimate $\sim 3000$~km~s$^{-1}$. Below, we discuss mostly velocity estimates involving proper motion measurements. 

In Table~\ref{tab:SNR}, we present measurements of 2D (projected) velocities of NSs in SNRs. Several values are taken from the compilation by \cite{2017ApJ...844...84H}, so we do not provide references to the original papers. 
Many measurements for CCOs were made recently by \cite{2021A&A...651A..40M}.
 Details, like accounting for the Galactic rotation, etc., may be found in the original papers. We just specify the cases where associations of an NS with an SNR are doubtful and mark if the velocity was obtained without a proper motion measurement.

\begin{longtable}[t]{l|c|c|c|l}
    \caption{Projected (2D) velocities of NSs in SNRs
    \label{tab:SNR}}\\
  Object       &  Type & Remnant & Velocity,  & Ref. \\
   &  & & km~s$^{-1}$ & \\
  \\
  \endfirsthead
  \multicolumn{5}{c}{Continuation of Table \ref{tab:SNR}}\\
    Object       &  Type & Remnant & Velocity,  & Ref. \\
       &  & & km~s$^{-1}$ & \\
    \\
  \endhead
  \\
\multicolumn{5}{l}{$^a$ Association of the NS with the remnant is doubtful.} \\
\multicolumn{5}{l}{$^b$ Velocity measured without the proper motion.}\\
    \endfoot
    \\
\multicolumn{5}{l}{$^c$ The authors claimed $v=180 (d/2\, \text{kpc})$~km~s$^{-1}$.}\\
\multicolumn{5}{l}{        The distance is estimated now as 3.3 kpc \citep{2021ApJ...922..253M}.}\\
\multicolumn{5}{l}{$^d$ The authors use two distance values. }\\
\multicolumn{5}{l}{The velocity given in the Table corresponds to the distance 3.8~kpc. }\\
\multicolumn{5}{l}{For the less favored value of 2.1 kpc, the transverse velocity is $264\pm 78$~km~s$^{-1}$. }\\
        \endlastfoot
  PSR J0007+7303 & PSR & CTA 1 & 450$^b$ & \cite{2017ApJ...844...84H} \\ 
  PSR J0205+6449        & PSR & 3C 58 & $35 \pm 6$ & \cite{2013MNRAS.431.2590B}\\ 
  SGR 0526-66           & SGR & N49 & $1100\pm 50^b$ & \cite{2018ApJ...856...18K} \\ 
  PSR B0531+21    & PSR & Crab & $104^{+13}_{-11}$ &  \cite{2023ApJ...952..161L} \\ 
  (Crab) & & & & \\
  PSR J0538+2817        & PSR & S147 & $400^{+114}_{-73}$ & \cite{2007ApJ...654..487N}\\ 
  CXOU J061705.3 & X-ray,   & IC 443 & 160 & \cite{2017ApJ...844...84H} \\ 
  +222127 & PWN & & & \\
  PSR B0656 + 14        &  PSR & Monogem & $60\pm 7$ & \cite{2005MNRAS.360..974H}\\ 
  RX J0822-4300         & CCO & Puppis A  & $763\pm 73$ &  Mayer, Becker (2021) \\
  PSR B0833-45          & PSR & Vela & $61\pm 2$ & \cite{2003ApJ...596.1137D} \\ 
 CXOU J085201.4  & CCO & G266.1-1.2   & $<1400$ & Mayer, Becker (2021) \\ 
-461753& & (Vela Jr.)  & & \\
 PSR J0908–4913$^a$     & PSR & G320.4–1.0 & $670\pm 130 $ & \cite{2021MNRAS.507L..41J} \\ 
 CXOU J111148.6  & X-ray,  & G291.0-00.1 & 303 & \cite{2017ApJ...844...84H} \\ 
 -603926 & PWN & & & \\
 PSR J1124-5916         & PSR & G292.0+01.8 &  $612 \pm 152$&  \cite{2022ApJ...932..117L} \\ 
 1E 1207.4-5209         & CCO  & PKS 1209 &  $<180$ & Mayer, Becker (2021) \\ 
  & & -51/52 & & \\
 Calvera$^a$            & X-ray &  G118.4+37.0 & $\sim 300^c$ & \cite{2015ApJ...812...61H} \\  
 1E1547.0-5408          & AXP & G327.24-0.13 & $280\pm 120$ & \cite{2012ApJ...748L...1D} \\ 
---                     & X-ray,  & G326.3-01.8 & $720^{+290}_{-215}$ & \cite{2017ApJ...851..128T} \\ 
& PWN & & & \\
---                     & PWN & G327.1-1.1 & $\sim 400^b$ & \cite{2009ApJ...691..895T} \\ 
CXOU J160103.1   & CCO & G330.2+1.0 & $<230$ & Mayer, Becker (2021) \\ 
-513353& & & & \\
PSR J1631–4722 & PWN & G336.7+0.5 & $125-225^b$ & \cite{2025MNRAS.537.2868A}  \\
   1WGA J1713.4    & CCO & RX J1713.7& $<230$ & Mayer, Becker (2021) \\ 
   -3949 & & -3946 & & \\
 XMMU J172054.5  & CCO & G350.1-0.3 & 320$^{+210}_{-190}$ & Mayer, Becker (2021) \\ 
-372652 & & & & \\
PSR B1727-47$^a$        & PSR & G343.0-6.0 & $\sim 400$ & \cite{2019ApJ...877...78S} \\ 
PSR J1809-2332          & PSR & G7.5-1.7 & $231\pm 46$ & \cite{2012ApJ...755..151V} \\ 
PSR J1811-1926$^a$      & PSR & G011.2-00.3 & $<108^b$ & \cite{2017ApJ...844...84H} \\ 
CXOU J181852.0   & CCO & G15.9+0.2 & $<1200$ & Mayer, Becker (2021) \\ 
-150213& & & & \\
CXOU J182913.1& X-ray,  & G018.9-01.1 & $474\pm129^d$& \cite{2025arXiv250701084H} 
\\
-125113$^a$ & PWN & & & \\
1E 1841-045             & AXP & Kes 73 & 367 & \cite{2017ApJ...844...84H} \\ 
CXOU J185238.6   & CCO & Kes 79 & $<450$ & Mayer, Becker (2021)  \\ 
+004020 & & & & \\
PSR J1856+0113          & PSR & W44 & 370 & \cite{2017ApJ...844...84H} \\ 
4FGL J1903.8       & PWN & 3C 396 & 344$^b$ & \cite{2017ApJ...844...84H} \\ 
+0531  & & & & \\
3FGL J1923.2      & PWN & W51C & 360$^b$ & \cite{2017ApJ...844...84H} \\ 
+1408e  & & & & \\ 
SGR 1935+2154           & SGR & G57.2+08 & $97\pm 48$ & \cite{2022ApJ...926..121L} \\ 
PSR B1951 + 32          & PSR & CTB 80 & $273\pm 11$ & \cite{2007ApJ...660.1357N}\\ 
1E 2259+586             & AXP & CTB 109 & 157 & \cite{2017ApJ...844...84H} \\ 
CXOU J232327.9  & CCO & Cas A & $570 \pm 260$ & Mayer, Becker (2021)  \\ 
+584842 & & & & \\
PSR B2334 + 61          & PSR & G114.3 + 0.3 & $164$ & \cite{2005MNRAS.360..974H} \\ 
\end{longtable}















Detailed studies of NSs in SNRs can provide insight into the SN explosion mechanism. 
For example, the analyses by
\cite{2017ApJ...844...84H} 
and
\cite{2018ApJ...856...18K}
suggest that NS kicks might be due to asymmetric matter ejection instead of asymmetric neutrino emission.

Using a sample of 18 SNRs associated with NSs, \cite{2017ApJ...844...84H} analyzed how the morphology of SNRs is related to the direction of the kick velocity. In the cases of the most accurate data, the authors demonstrate that 
the direction toward the bulk of the X-ray emission is opposite to the direction of the kick. This indicates that NSs are kicked opposite to the bulk of the SN ejecta. 

\cite{2018ApJ...856...18K} presented a more detailed study of six SNRs in which they determined the directions of the ejecta of three groups of elements: iron, oxygen, and intermediate-mass elements that include Si, S, Ar, and Ca. 
It is shown that the centers of mass of the intermediate element ejecta are shifted from the explosion sites in the direction opposite to the motion of NSs in all remnants. In addition, it is shown that the kick velocity correlates with the degree of asymmetry of the ejection of the intermediate elements. This matches the predictions of hydrodynamical models of SN explosions. 

At the end of this subsection, we mention the very special case of SN 1987A. 
JWST spectral measurements probably allow for the estimation of the kick velocity of the compact object, most probably an NS \citep{2024Sci...383..898F, 2025arXiv250803395L}. 
The authors base their conclusions about the kick on the properties of the Ar~VI line emission. 
The origin of the emission is shifted from the center of the remnant. In addition, the line is blue-shifted. 
Together, this provides the value of the 3D kick velocity as $510\pm55$~km~s$^{-1}$. 
Unfortunately, at the moment, the spin properties of the compact object are not known definitively.  Thus, it is not possible to say much about the relative orientation of the velocity and spin axis. 
However, the kick direction appears to be perpendicular to the triple ring structure, suggesting that the kick was in the equatorial plane of the binary whose merger created the progenitor of 1987A.  Meanwhile, if the bipolar iron jet axis indicates the NS spin direction, this appears to be misaligned relative to both the progenitor spin and the kick \citep{2023ApJ...949L..27L}.
In the next subsection, we discuss the present evidence for the spin-velocity correlation.


\subsection{Kick/spin correlations}
\label{s:kick_spin}

A discussion of the relative directions of the spin axis and the spatial velocity vector began with theoretical considerations of an SN explosion. Early measurements of large spatial velocities of PSRs required some acceleration mechanism to explain them. Acceleration could happen rapidly due to the explosion (kick) or later during the NS evolution (Section~\ref{s:models}).\footnote{We do not discuss here kick/spin correlations in application to BHs. The interested reader can look at \cite{2024arXiv241203461B} and references therein. This topic is particularly important in application to gravitational wave data on BH coalescences, as the spin-orbit misalignment can be inferred from observations.} 
Different acceleration mechanisms predict various relative orientations of the spin axis and the spatial velocity. For example, the electromagnetic rocket predicts spin-velocity alignment, which is more pronounced for older pulsars than for very young ones \citep{1977ApJ...216..842H}. By contrast, the formation of an ultra-compact NS-NS binary due to an instability of the proto-NS might result in a kick perpendicular to the spin axis \citep{2002ApJ...581.1271C}. 
Thus, it is necessary to probe observationally if there is any correlation between the kick and spin directions.

From an observational point of view, the situation remained unclear until the 21st century, as early studies were contradictory and inconclusive. 
The first robust evidence for the spin-velocity alignment (confirmed by later studies) was proposed by \cite{1999A&A...344..367C} 
for the Crab and by \cite{2001ApJ...556..380H} 
for the Vela pulsar. 
Both results are mainly based on the symmetry of X-ray structures around the pulsars: the symmetry axis, which is assumed to be the spin axis, is co-aligned with the proper motion direction. 
Later on, PSR J0538-2817 was added to the list using the same technique
\citep{2003ApJ...585L..41R}. 
The alignment was demonstrated by \cite{2004ApJ...601..479N} for a larger set of six PSRs in PWN. 
Finally, \cite{2021NatAs...5..788Y} 
were able to demonstrate that the spin axis and velocity vector of PSR J0538+2817 are co-aligned within a few degrees. This became possible thanks to measurements of the radial velocity of the PSR; however, the uncertainties are significant. 
 Still, in order to confirm the general tendency for alignment, it is necessary to use larger samples. To reach this goal, one might include pulsars without a structured PWN.

 Proper motions are measured for many pulsars. Determining the orientation of the spin axis is a more challenging task. This can be done with polarimetric measurements. When the radio emission is linearly polarized, it is observed that the polarization angle, PA, changes with the spin phase often showing an S-shaped curve. In the simplest case of the parallel polarization mode, PA rapidly changes when the spin axis, the magnetic dipole axis, and the line of sight are in one plane, i.e., when the line of sight is in the closest position with respect to the pulsar's beam. 
 Neglecting some complications related to the emission mechanism and propagation of the radio waves, it is often assumed that the PA at the center of the S-shaped curve corresponds to the orientation of the spin axis. Even if the S-shaped curve is not observed, in many cases it is still possible to obtain an estimate of the spin axis orientation. 

 As radio waves can be emitted in the so-called orthogonal polarization mode, there is a 90-degree ambiguity in the position of the spin axis, see, e.g., \cite{2007ApJ...664..443R}. In this case, the emission is polarized perpendicular to the magnetic field lines.
 For example, for the Vela pulsar, the spin direction is determined independently using X-ray data, and it is possible to demonstrate that the polarization measurements define the direction perpendicular to the spin axis orientation, see, e.g., Section~6.2.1 in \cite{2005MNRAS.364.1397J} and references therein. Thus, accounting for strong arguments in favour of spin-velocity alignment, it is sometimes assumed that the angle between the velocity vector and the direction derived from the polarization measurements is $<45^\circ$. I.e., if the measured angle under the assumption of the parallel polarization mode appears to be $\mathrm{PA}_\mathrm{meas}>45^\circ$, 
 then it is assumed that the emission is in the orthogonal mode and the spin axis orientation corresponds to $90^\circ - \mathrm{PA}_\mathrm{meas}$, see, e.g., \cite{2007ApJ...660.1357N, 2012MNRAS.423.2736N}. 

 One of the first large samples of PSRs with reliable (according to modern standards) proper motion and polarization properties was presented by
\cite{2005MNRAS.364.1397J}.  
The authors studied 25 pulsars. For 10 of them, they found that the spin-velocity angle is $<10^\circ$ or $>80^\circ$ (due to uncertainty in the polarization mode). For another 10 PSRs, there is no clear relation between spin and velocity directions. Finally, for 5 PSRs polarization measurements did not allow determining the spin axis orientation. 
Young PSRs show better spin-velocity correlation. These results allowed \cite{2005MNRAS.364.1397J} to conclude that accounting for the parallel/orthogonal polarization ambiguity, spin-velocity alignment is strongly favoured. 

The results of \cite{2005MNRAS.364.1397J} were supported and strengthened by further studies \citep{2007ApJ...664..443R, 2007MNRAS.381.1625J}.
A large sample of PSRs with well-measured or constrained kick-velocity angles was used by \cite{2007ApJ...660.1357N} 
 to test predictions from several models of the kick generation. 

 The spin axis direction can be determined from a polarization measurement if the PSR emission demonstrates robust features of the central core-beam.
 One of the largest (up-to-date) sets of such PSRs with well-measured proper motions and good polarization data is presented by \cite{2015ApJ...804..112R}, who analyzed 47 objects. The author concluded that the core emission is mostly in the orthogonal polarization mode, and there is a very strong tendency toward spin--velocity alignment. 



Another large set -- 54 PSRs -- was studied by \cite{2012MNRAS.423.2736N}. 
This sample is dominated by polarization measurements (51/54), but three PSRs with X-ray tori observed by {\it Chandra} are also included (J0534+2200, J1709-4429, J1952+3252). About 70\% of the objects have an angle between the spin and velocity axes of $<20^\circ$. This excludes the absence of correlation between these directions at a $>3\sigma$ level. 

It is important that any correlation between spin and velocity directions might be lost on a time scale $\gtrsim 10^7$~yrs due to the influence of the Galactic gravitational potential. This effect has been studied by
\cite{2013MNRAS.430.2281N}. 
The authors used spin-down and kinematic ages. 
Spin-down or characteristic ages can be significantly different from the true ages for many reasons: the present-day spin period may be close to the initial one, the magnetic field may decay, etc. Thus, it is not surprising that the analysis of spin-velocity alignment for PSRs with different characteristic ages does not demonstrate blurring of the correlation with age. 
Then, \cite{2013MNRAS.430.2281N} derived reliable kinematic ages for 33 of the 58 PSRs under study. In this case, the distribution of the angle between the spin axis and velocity demonstrates flattening for ages $\gtrsim 10$~Myr, in correspondence with expectations.  

Note that even if a kick mechanism produces significant alignment (of course, different types of SN can result in different relative orientations of spin and velocity axes), not all young PSRs might demonstrate it, as some of them may originate in close binary systems. In that case, the pre-supernova orbital velocity might non-trivially contribute to the total observed velocity. The influence of binarity of progenitors was studied, e.g., by \cite{2009MNRAS.395.2087K}. 
The authors performed a population synthesis for binary and isolated NS progenitors for various kick parameters. The main goal was to derive the range of observable parameters (velocity and alignment angle) for which binary vs.~isolated progenitors dominated. \cite{2009MNRAS.395.2087K} demonstrate that excluding low-velocity NSs, isolated (and wide binary) progenitors dominate the population with tight alignment ($<10^\circ$). In general, the contribution of binary companions is important to explain the observed alignment distribution. Still, the present-day statistics of alignment and poor understanding of the initial parameters prevent detailed conclusions (especially for low-velocity NSs).

If the spin axes of young PSRs are predominantly directed close to their velocity vectors, then an interesting selection effect might appear.
\cite{2023ApJ...944..153M} 
noticed that in the case of spin-velocity alignment, the measured 2D velocity of observable pulsars is not an
isotropic projection because the detectability of a
 pulsar depends on the location of the observer relative
 to the spin axis. Thus, when one reconstructs the 3D velocity from the proper motion measurements, this effect must be taken into account. 
This approach was applied to several dozen radio pulsars by 
 \cite{2024arXiv241212017B}. These authors demonstrated that peculiar velocities of weakly oblique and strongly oblique pulsars are distributed differently. Their analysis supports the idea of the spin-kick alignment. 

 The effect depends on the angle between the magnetic and spin axes of the PSR. 
 If this angle is small, then the 3D velocity is significantly larger than the 2D one. By contrast, if the angle is close to $90^\circ$ then the 3D velocity is nearly equal to the measured 2D value. Thus, it is necessary to make assumptions about the distribution (and evolution) of this angle. An additional dependence appears on the width of the PSR's beam. All these parameters can evolve over time. Thus, an accurate account of the effect is rather complicated. However, it can influence the determination of 3D velocities of young ($\lesssim 10$~Myr) PSRs at the level of tens of percent. 
 
In the previous subsection, we mentioned several results related to correlations of SNR properties with the NS velocity direction. At the end of this subsection, it is worth mentioning another result of a similar kind.
\cite{2018ApJ...855...82B}  
studied how the velocity direction is related to another geometrical feature of SNRs, which the authors attribute to the jet produced during the SN explosion  (see \citealt{2025RAA....25d5008B} for a discussion of the proposed mechanism kicks by asymmetric jet emission). 

The idea is as follows. If powerful jets have been generated, then they might produce morphological features in the remnant (e.g., `ears' or bright arcs). The line connecting these features defines the jet orientation, which can be compared to the orientation of the spatial velocity of the NS (if its proper motion is measured). Thus, one can determine how the jet direction is correlated with the velocity vector. Using 12 NS-SNR pairs, the authors demonstrate that the relative orientation of the two directions is almost consistent with a random distribution. Just a slight shortage of small angles ($\lesssim 40^\circ$) was identified, i.e., there is no clear argument in favor of jet-kick alignment (however, in the specific case of the NS in SNR S147, \cite{2025arXiv250621548S} came to a different conclusion that jets are responsible for the kick). If this result can be confirmed with larger statistics  and with a less subjective method, it might be very important for understanding NS acceleration, as well as for understanding the SN mechanism. Note, however, that the interpretation of features in the supernova remnants as ears or jets is not stringent and remains open to debate.


It is not clear how to interpret the spin-velocity alignment. Three basic possibilities are the following:

\begin{itemize}
\item averaging along the spin direction if the kick mechanism operates over many spin periods;
\item the kick direction is determined by the spin direction;
\item the spin is obtained as a consequence of the kick.
\end{itemize}

If a randomly (but not isotropically) directed kick mechanism is relatively long (at least several spin periods), then the imparted momentum is averaged along the spin axis. However, if the accelerating mechanism works for many spin periods, then one expects a perfect alignment (like in the electromagnetic rocket mechanism), which is not observed. Otherwise, the kick direction can be determined explicitly by the spin direction, i.e., the asymmetry of the accelerating mechanism is strongly spin-dependent. This can happen, for example, if the kick mechanism is related to the toroidal magnetic field amplified in the equatorial plane, see e.g., \cite{2024PhRvD.110h3025K} and references therein. Finally, an NS can obtain spin due to the kick mechanism. Irrespective of the mechanism of momentum gaining, the velocity vector orientation can determine the direction of the angular momentum increase producing alignment. 
For example, an NS can be spun up by fallback accretion of matter with significant angular momentum perpendicular to the velocity vector \citep{2022ApJ...926....9J}, 
see, however, \cite{2023MNRAS.526.2880M} for a counter-argument. 

\subsection{Microlensing}

NSs and BHs are not very numerous in comparison to normal stars or even WDs.  NSs and BHs constitute a few tenths of one percent and $\sim 0.1\%$ of the present-day Galactic stellar population by number, respectively  (e.g., \citealt{2010A&A...523A..33S}). 
However, as NSs are $\approx 3-5$ times more massive than an average star and BHs are $\approx 3-10$ times more massive than NSs, the probability of lensing by these sources is non-negligible because the Einstein radius grows as $\sqrt{M_\mathrm{lens}}$, where $M_\mathrm{lens}$ is the mass of the lensing object. 
It is expected that significant statistics of microlensing events by isolated compact objects might provide a reliable mass distribution for them \citep{1996MNRAS.278L..46H, 2000ApJ...535..928G}. However, degeneracies involving the spatial velocities of NSs and BHs play an important role.

 Curiously, one of the first papers related to gravitational microlensing by NSs \citep{1997ApJ...479..147M} already discussed the issue of their high velocities. These authors suggested that due to large kicks, NSs can contribute significantly to the rate of microlensing in Large Magellanic Cloud observations. 

 The rate of lensing by NSs and BHs has been modeled in many papers since 2000. We do not intend to give a full list, just to mention a few: \cite{2000ApJ...535..928G, 2002A&A...388..483S, 2008A&A...478..429O, 2010A&A...523A..33S, 2024MNRAS.531.2433S}. It is estimated that a few per cent of all microlensing events are due to either NSs or BHs, but for long-duration events ($\gtrsim 100$ days), their contribution rises to a few tens of percent. 
 Calculations of the rate of microlensing events explicitly include assumptions about kick velocity. In the first place, velocity plays a role in modelling the trajectories and the spatial distribution of compact objects in the Galaxy. Then, the value of the tangential velocity of the lens relative to the source and observer determines the duration of the event. Finally, the probability of lensing depends on the relative velocity as well, as a lens with a higher velocity covers a larger patch of the sky during the time of the survey. 

 Proper motions of high-velocity or/and nearby known NSs can be measured with high precision. 
 This gives an opportunity to check if they can serve as gravitational lenses for known sources. 
 E.g., this was done for one of  the nearby cooling NSs (so-called Magnificent Seven, see \citealt{2023Univ....9..273P}) RX J1856-3754, see \cite{10.1093/mnras/stx2985} and references therein. Also, such a possibility has been studied for many radio pulsars, e.g. \cite{2002A&A...388..483S}, but with a negative result. Still, with increasing number of pulsars (mainly due to FAST and future SKA observations), it might be possible to observe microlensing events with known pulsar lenses \citep{2010MNRAS.405.2754D}.

 During the last few years, many papers related to BH microlensing (and its connection to the kick velocity) have been published. Currently, there are many candidate photometric BH events \citep{2016MNRAS.458.3012W, 2025A&A...694A..94H, 2025ApJ...981..183K} and one well-established case for which both photometric and astrometric data are available \citep{2025ApJ...983..104S}. Population synthesis modelling of BH microlensing was presented, e.g., by \cite{2020ApJ...889...31L}. 

 OGLE-2011-BLG-0462 is the most reliable event classified as microlensing by a BH \citep{2025ApJ...983..104S}.
 The mass of the lens is $7.15\pm 0.83\, \mathrm{M}_\odot$. What is more important for our discussion is that the velocity of the compact object is measured, too. The transverse velocity is $51.1 \pm 7.5 $~km~s$^{-1}$ relative to its local standard of rest. This is significantly larger than the velocities of stars near the lens. Thus, the BH might have obtained a kick at its birth. In the future, the number of well-studied BH events will increase. The Nancy Grace Roman space telescope is expected to contribute significantly \citep{2024arXiv240706484G}. In particular, \cite{2020ApJ...889...31L} predict that this instrument can obtain a few hundred mass measurements for BH lenses. Then it might also be possible to obtain the velocity distribution of isolated BHs from observations. 

 Still, even one well-studied case of BH microlensing allowed for some constraints on the kick velocity properties. For example, \cite{2022ApJ...930..159A} demonstrated that despite all uncertainties (the BH could be born in a thick or thin disc, the progenitor could be a single star or a member of a binary), most likely the BH received a kick at birth and its value is $\lesssim 100$~km~s$^{-1}$. 
A similar conclusion is reached by \cite{2024ApJ...973....5K} who used a completely different approach and calculated the probability of observing an event similar to OGLE-2011-BLG-0462. These authors demonstrate that the observed BH lens might come from a population with an average kick velocity distribution $\lesssim 100$~km~s$^{-1}$. 

Other indirect evidence exists in favor of modest kicks of BHs.
\cite{2020A&A...636A..20W} 
demonstrated\footnote{See some critiques and discussion in \cite{2021AcA....71...89M}.} that if a mass gap in the range 2-5~$\mathrm{M}_\odot$ is assumed for isolated BHs,  then kicks of a few tens of km~s$^{-1}$ are necessary to explain OGLE-III data. However, due to the degeneracy between transverse velocity and the mass of the lens,  photometric data alone  is not sufficient to put assumption-independent constraints on the kick. The typical timescale of the event is:

\begin{equation}
t_\mathrm{E}\propto \theta_\mathrm{E}/\mu_\mathrm{rel}.
\end{equation}
Here, $\theta_\mathrm{E}$ is the Einstein radius 
and $\mu_\mathrm{rel}$ is the relative proper motion of the lens and the source. Thus, for a given measured $t_\mathrm{E}$ a larger transverse velocity
can compensate for a larger mass of the lens. Joint astrometric and photometric measurements can remove this degeneracy. 




\subsection{Detached binaries: eccentricities and other properties}



Recently, about two dozen NS and BH binary systems of a new type have been reported, see
\cite{2024OJAp....7E..58E}, 
\cite{2024A&A...686A.299S} 
and references therein. 
These are wide non-interacting binaries in which a NS or a BH has a solar-like star as a companion.
Typical orbital periods are about 100-1000 days. Orbital eccentricities are in the range 0.1-0.8 with a median value $\sim 0.4$. The systems have been discovered using Gaia satellite observations. 

The present-day total masses of such detached NS binaries are approximately $2-2.5\, \mathrm{M}_\odot$. Thus, if before the SN explosion the NS progenitor had a mass $\gtrsim 4\, \mathrm{M}_\odot$, then for a zero kick the system might become unbound as more than half of the total mass is instantly lost. However, even in the case of a massive progenitor, the system can remain bound if an appropriate kick is received. \cite{2024OJAp....7E..58E} analyzed this possibility for several progenitor masses.

As presently the systems have orbital periods of about a year, we can be sure that the two stars have interacted with each other before the supernova. For example, the systems could have experienced dynamically unstable mass transfer in a common envelope episode, see reviews by \cite{2013A&ARv..21...59I, 2020rfma.book..123J, Ivanova:2020}. 
Then the NS progenitor may have already have lost its hydrogen envelope before the SN explosion. 
Thus, \cite{2024OJAp....7E..58E} considered helium stars with masses 3, 4, 5, and 10 $\mathrm{M}_\odot$ as progenitors and estimated the probability of a system to survive for various values of the kick velocity isotropically imparted to the NS. 
 Note that in their analysis, \cite{2024OJAp....7E..58E} did not use information on the velocity of the center of mass of the binary system. Potentially, such information, together with a measured eccentricity for wide systems where tidal forces cannot significantly circularize the orbit, might provide an additional constraint on the parameters, eliminating degeneracy between kick and mass loss. 

Only in the case of the $3\, \mathrm{M}_\odot$ progenitor, the probability to survive is maximal for the zero kick. 
In other cases, moderate kicks $\sim (10-50)$~km~s$^{-1}$ are preferred, depending on the progenitor mass and orbital period prior to the SN explosion. For kicks $\gtrsim 100$~km~s$^{-1}$, the systems are mostly  disrupted. 
The existence of wide NS binaries with low-mass companions confirms that NSs can experience low kicks similar to those needed to explain the NS population in globular clusters.

Some of the wide low-mass binaries with NSs have low eccentricities e.g., Gaia~NS1 with $P_\mathrm{orb}=731$~d and $e\approx0.12$ \citep{2024OJAp....7E..27E}. It is nontrivial to explain such properties within the standard framework of binary evolution. 
Usually, only instantaneous kicks are considered in binary system modeling. However, as discussed above in Section~2, NS acceleration can happen on a longer time scale, e.g., due to the electromagnetic rocket mechanism. 
If the time scale of acceleration is significantly longer than the orbital period, then its influence on the orbital parameters must be considered appropriately.
This has been done by
\cite{2024ApJ...972L..18H, 2025arXiv250705602B}.

These authors demonstrated that the addition of electromagnetic rocket acceleration can decrease the eccentricity of the system. Thus, for Gaia NS1, they propose the following scenario.  (i) An initially wide system with the semimajor axis $\lesssim$~a few AU enters  (ii) the common envelope stage. The orbit shrinks down to an orbital period of $\lesssim$~a few days. (iii) The NS progenitor loses its hydrogen envelope and explodes in an SN. Due to mass loss and a natal kick, the system becomes wide (with orbital period $\gtrsim 100$~d) and highly eccentric.  (iv) Afterwards, due to the electromagnetic rocket mechanism, the eccentricity diminishes down to the value we observe today.

Interestingly, the symbiotic X-ray binary IGR J16194-2810 has properties similar to those of Gaia NS1 
 \citep{2024PASP..136g4202N}. 
It has $P_\mathrm{orb}\sim 200$~d and $e\approx 0$. There also a few other similar systems with $P_\mathrm{orb}\sim (100-1000)$~d and $e\approx (0.1-0.3)$, see, e.g., Fig.~5 in \cite{2024ApJ...972L..18H} and Table 1 in \cite{2024PASP..136g4202N}. 

\cite{2024PASP..136g4202N} modelled binary evolution, which can lead to the formation of systems similar to the observed one. 
Assuming that the present-day low eccentricity is due to tidal interaction between the giant donor and the NS, for He-star progenitors with masses $2-5\, \mathrm{M}_\odot$ the system can be reproduced for a wide range of initial orbital periods $\approx 50-400$~d and kicks $\lesssim 50 $~km~s$^{-1}$. The authors suggest that symbiotic X-ray binaries with NSs are mostly descendants of systems such as the wide low-mass binaries discovered by Gaia. With an increased number of such systems -- which is expected as soon as Gaia DR4 will become public,  see \cite{2025arXiv250821805C}  -- and detailed population synthesis modelling, it would be possible to put important constraints on the low-velocity part of the NS kick distribution.

 New discoveries of wide non-interacting binary systems with compact objects provide an opportunity to also constrain the kick velocity distribution of BHs. 
 Currently, three systems with non-accreting BHs and detached companions have been discovered by Gaia \citep{2024NewAR..9801694E}.
\cite{2024MNRAS.535.3577K} 
 focused on Gaia BH1 and Gaia BH2. 
These systems have $P_\textrm{orb}=191$~d and 1592~d, respectively. Companions are solar-like stars. 
The orbits are moderately eccentric, with eccentricities of 0.48 and 0.56. 
Both systems  analyzed by \cite{2024MNRAS.535.3577K}  can be successfully reproduced in the standard scenario of binary evolution. In the case of Gaia BH1, the kick velocity might be $\lesssim 100$~km~s$^{-1}$ and in the case of Gaia BH2 $\lesssim 50$~km~s$^{-1}$. Such velocities naturally come out when assuming the bimodal kick distribution of \cite{2023MNRAS.525.1498Z} reduced due to the fallback. The same calculation assuming the velocity distribution proposed by \cite{2005MNRAS.360..974H} (also reduced in the case of BHs) yields a reduced formation rate, which is not sufficient to explain the observed systems \citep{2024MNRAS.535.3577K}. 
Similar results from an independent analysis can also be found in the paper by \cite{2025PASP..137c4203N}. 

NS low-mass wide binary systems discovered by Gaia are likely to be old systems. In their observed state, they can live for $\sim10$~Gyr, the lifetime of the low-mass companion on the main sequence. For a short fraction of this time, $\sim 10$~Myr, a young binary of this type may contain an active radio pulsar. Such systems are rare. The ANTF catalogue \citep{2005AJ....129.1993M} does not contain any robust examples of them with identified companions.
Still, many interesting binaries with radio pulsars are known, and now we briefly discuss one of them, containing a main-sequence companion.

 PSR J0210+5845 is observed in an extremely wide binary with a main sequence companion identified as a B6V star with mass 3.5-2.8$\mathrm{M}_\odot$ \citep{2024A&A...682A.178V}.
 The orbital period is estimated as $47^{+40}_{-14}$~yrs.  The orbital eccentricity is $0.46^{+0.10}_{-0.07}$. 

 As in the case of Gaia binaries (see \citealt{2024OJAp....7E..58E} and beginning of this subsection), if the progenitor of the pulsar was a low-mass helium star, then very low kicks (with that maximum at zero) are favoured. If the progenitor had $M=8-10 \, \mathrm{M}_\odot$ before the explosion then kicks are expected in the range $\sim10 - 40$~km~s$^{-1}$ and for kicks $\gtrsim 100$~km~s$^{-1}$ the system should become unbound \citep{2024A&A...682A.178V}. These results led the authors to the conclusion that the NS was born in an electron-capture SN. 

The properties of pulsars born in an electron-capture SN in very wide systems have been studied by \cite{2022MNRAS.513.6105S}. They performed a simplified modelling of very wide non-interacting binary systems. The minimum initial orbital period was taken to be $10^4$~days. It was assumed that electron-capture SNe happen in a narrow range of progenitor masses with a width of $0.2\, \mathrm{M}_\odot$ around $8\, \mathrm{M}_\odot$ and provide kicks  $\sim$10~km~s$^{-1}$. Due to the assumed power-law distribution of the mass ratio with a slope $-2$, most companions are low-mass stars. 
The formation rate of radio pulsars formed through an ECSN in surviving wide binaries is about one per 1000 isolated pulsars. 
However, those with $P_\mathrm{orb}\gtrsim100$~yrs might be misclassified as isolated sources, see e.g. \cite{2023ApJ...951...20J}. The pulsars with tighter orbits form a minority, and their fraction is an order of magnitude lower, i.e., approximately one pulsar formed through an ECSN might be observed in a wide binary for every 10,000 observed isolated pulsars. 

Constraints on the kick velocity distribution may come from  disrupted binaries if both components are identified, not just extant binaries.
Suppose that we observe an isolated NS with a measured velocity. With a significant probability, this NS was born in a binary (or a triple) system  disrupted by the SN explosion, see, e.g., 
\cite{2021MNRAS.507.5832K, 2024MNRAS.535.1315B}. 
Then, the present-day velocity is a combination of the kick and the orbital velocity (which is unknown). In addition, the age of the NS is not well-known, and thus it is difficult to account for the velocity evolution in the Galactic potential. However, if the NS is identified as a member of a  disrupted binary (i.e. if the secondary component is found) and the mass and velocity of the second star are measured, then it is possible to put strict constraints on the value of the kick. 

\cite{2024MNRAS.531.2379S} 
searched for possible companions (bound or unbound) for the known magnetars.  In the context of our discussion, we mention just two results of that paper. Firstly, the finding that the magnetars 3XMM J185246.6+003317 and CXOU J185238.6+004020 could originate from the same binary provides an estimate of their kicks at $\sim 300$~km~s$^{-1}$.  Secondly, the analysis of the binarity of magnetars drives the authors to the conclusion that low-kick ECSNe might not produce magnetars, as otherwise the number of these sources in bound binaries would be higher, in contradiction with observations.


\subsection{Magnetar kicks} 

 Historically, velocity measurements for magnetars have been of special interest. If it could be shown that these high-magnetic-field NSs have a notably different velocity distribution, then this would be an argument in favour of a specific origin (e.g., a different kind of SN explosion). Initially, it was suggested that magnetars have much higher velocities than other NSs \citep{1994ApJ...431L..35V}, see also Table 1 in \cite{2017MNRAS.464.4895G}.
 Later, it appeared that the early velocity determinations were not correct.

 Now, the situation seems to be opposite. \cite{2024ApJ...971L..13D} demonstrated that the average velocity of magnetars is similar to or smaller than the average velocity of young radio pulsars. Using a set of 8 magnetars these authors estimated the average 3D velocity as $190^{+168}_{-87}$~km~s$^{-1}$. 
 Recently, the velocity measurement for the 9th member of the magnetar family was reported \citep{2025A&A...696A.127C}. This is the source SGR 0501+4516. Its transverse velocity is even lower the average of the other 8 magnetars: $51\pm 14$~km~s$^{-1}$. 

 The precision of the transverse velocity measurements for magnetars suffers from uncertainties in distances, which in some cases are large, see, e.g., \cite{2022ApJ...926..121L}. The systematic distance uncertainty is not always included in the uncertainty in the transverse velocity. However, most well-studied magnetars have measured velocities $\lesssim 200$~km~s$^{-1}$. 

 As magnetars are very young objects with typical ages $\lesssim 10^4-10^5$~yrs \citep[e.g.,][]{2015RPPh...78k6901T}, selection effects against high velocity NSs, which can leave the area of detectability, are not applicable. 
 Having $\sim 30$ magnetars \citep{2014ApJS..212....6O} for nine of which transverse velocities are measured, 
 we can conclude that in general, magnetars are moderate-velocity objects. Their velocity distribution is not consistent with the one derived by \cite{2005MNRAS.360..974H} for radio pulsars (but see above regarding a correction to this fit). However, it can be consistent with the low-velocity part of bimodal distributions, e.g. those proposed by \cite{2017A&A...608A..57V} and \cite{2020MNRAS.494.3663I}, see \cite{2024ApJ...971L..13D} for discussion, or with the single log-normal distribution of \citet{2025arXiv250522102D}.

\begin{table}[t]
    \centering
    \begin{tabular}{l|c|c|c|c}
  Object       &  Distance,  & Band  & Velocity,  & Ref. \\
  & kpc & & km~s$^{-1}$ & \\
  & & & & \\
4U 0162+61     & $3.6\pm 0.4$  &  NIR & $102\pm 26$  & \cite{2013ApJ...772...31T} \\ 
SGR 0501+4516  & 2 &  NIR &   $51\pm 14$ &  Chrimes et al. (2025)  \\ 
1E 1547.0-5408 &  $6\pm 2$ & Radio & $280^{+130}_{-120}$  & \cite{2012ApJ...748L...1D}\\ 
SGR 1806-20    & $9\pm 2$ & NIR & $350\pm 100$  & \cite{2012ApJ...761...76T}\\ 
XTE 1810-197$^a$   & $2.5^{+0.4}_{-0.3}$ &  Radio &   $198^{+29}_{-23}$ & \cite{2020MNRAS.498.3736D}\\ 
Swift J1818.0-1607$^a$ &  $9.4^{+2.0}_{-1.6}$  &  Radio &  $48^{+50}_{-16}$   & \cite{2024ApJ...971L..13D} \\ 
SGR 1900+14    & $12.5\pm 1.7$ & NIR &   $130\pm 30$ & \cite{2012ApJ...761...76T}\\ 
SGR 1935+2154  & $6.6\pm 0.7$ & NIR &  $97\pm 48$ & \cite{2022ApJ...926..121L} \\ 
1E 2259+583    &  $3.2\pm 0.2$ & NIR & $157\pm 17$  & \cite{2013ApJ...772...31T}\\ 
\end{tabular}
    $^a$ Distance is obtained via parallax measurements.
    \caption{ Transverse velocities of magnetars}
    \label{tab:Magnetars}
\end{table}
 
\section{Indirect observational constraints on kicks}

\label{s:indirect}

\subsection{Retention in globular clusters}

Nearly 350 pulsars are known to exist in Galactic globular clusters\footnote{\url{https://www3.mpifr-bonn.mpg.de/staff/pfreire/GCpsr.html}}.  These clusters have escape velocities of only a few tens of km s$^{-1}$, almost always less than 50 km s$^{-1}$.  This has presented a challenge for at least the past 30 years: only a few percent of single pulsar velocities fall below 50 km s$^{-1}$ and could thus be retained in globular clusters \citep{Drukier:1996}.  Pulsars born in binaries could have a higher chance of being retained due to the extra inertia provided by the binary companions.  However, typical single-pulsar kicks would disrupt most wide binaries, and the overwhelming majority of NSs would still escape globular clusters, which appears to be inconsistent with the large observed populations once observational selection effects are accounted for \citep{Pfahl:2002}.  

This suggests that there must be a significant low-kick channel that is not observed in the single-pulsar population.  One possibility, mentioned in Section \ref{s:pulsars}, is that low natal kicks are preferentially received in tight interacting binaries, with the resulting pulsars remaining locked up in such binaries and thus missing from the isolated-pulsar population. One example could be electron capture supernovae whose progenitors transferred mass to their companions \citep{Willcox:2021}.  Another could be white dwarf accretors becoming NSs through accretion induced collapse.  

It should be noted that NSs in globular clusters have an increased probability of being recycled through dynamical interactions, which enhances their long-term detectability \citep{SigurdssonPhinney:1995}. However, the actual number of NSs in globular clusters is not known, with large error bars on the number of undetected objects due to the unknown luminosity function \citep{Chennamangalam:2012}.  Meanwhile, some NSs may be ejected through subsequent dynamical interactions after the initial natal kicks.   It is therefore not trivial to accurately determine the actual fraction of retained NSs in globular clusters and to use it as a precise constraint on the kick distribution.

Retention of NSs and BHs in ultra-compact dwarf galaxies provides an additional constraint on natal kick velocities \citep{Pavlik:2018}.

\subsection{Systemic velocities of  compact-object binaries}

The vast majority of massive stars are born with companions: in binaries ($\approx 30\%$), triples ($\approx 30\%$) and quadruples or higher-order systems ($\approx 20\%$) \citep{MoeStefano2017ApJS}.  Approximately $70\%$ of massive binaries interact and exchange mass during their evolution \citep{Sana2012Sci}.  Thus, it seems that binaries hold the key to measuring compact-object velocities.

Indeed, a significant amount of effort has gone into measuring the velocities of compact-object binaries.  These include NSs in binaries of various flavours: low-mass X-ray binaries, red-back and black-widow pulsars, and millisecond pulsars, as well as BH X-ray binaries.  Historically, most of these observations included relatively accurate measurements of angular position in the Galaxy and the binary's systemic proper motion, but, as with single pulsars, struggled with accurate measurements of distance to the source and radial velocity.  

In a series of papers, Repetto and collaborators investigated the systemic velocities that BH binaries must have received in order to rise to their observed Galactic latitudes by considering the difference in the gravitational potential energy between the observed location of the binary and its projection onto the Galactic plane.  They found that at least some BHs must have received very significant kicks of several hundred km s$^{-1}$ \citep{Repetto2012MNRAS}.  However, these conclusions are sensitive to the large distance uncertainties and potentially to some of the assumptions in their models \citep{Mandel:2015kicks}, though see \citet{Repetto2017MNRAS}.

\citet{2019MNRAS.489.3116A} took a different approach to the unknown radial velocities of BH binaries by integrating their orbits backward in the Galactic potential and considering the kick velocities they must have received at Galactic plane crossings, under the assumption that the progenitor massive stars are born in a thin disk (and not, say, ejected from globular clusters).  The set of potential velocities at Galactic plane crossings was then interpreted as draws from the posterior on the binary systemic kicks.  They found typical potential velocities, i.e., binary systemic velocities, of $\sim 100$ km s$^{-1}$.

\citet{ODoherty:2023} applied the \citet{2019MNRAS.489.3116A} technique to a total of 145 NS binaries, including low-mass X-ray binaries, spiders, and millisecond pulsars, using parallax measurements, optical companion lightcurve fitting, and dispersion measures to estimate distances.  They found typical systemic velocities of $\sim 140$ km s$^{-1}$ for NS low-mass X-ray binaries and spider binaries, but only half that for millisecond pulsars. \citet{2021MNRAS.508.3345I} and \citet{Fortin:2022} explored NS high-mass X-ray binaries, particularly Be X-ray binaries.  

Recently, astrometric measurements of detached NS and BH binaries have become possible.  These include several Gaia detached BHs \citep{ElBadry:2023b,ElBadry:2023a}.  At least one detached BH binary appears to be kinematically associated with a stellar stream of a disrupted cluster \citep{Panuzzo:2024}, pointing to one of the challenges of using systemic velocities to make inferences about natal kicks.  Meanwhile, more than 20 detached NS binary candidates were identified in Gaia data \citep{2024OJAp....7E..58E}, as discussed in the previous section.

Unfortunately, even when accurate systemic velocities are available, inference on the compact object's natal kick relies on imperfectly understood models of binary evolution.   The systemic velocity depends on the properties of the binary system at the time of the supernova.  The \citet{1961BAN....15..265B} kick from symmetric mass loss depends on the pre-supernova orbital velocity and the fraction of mass lost in the supernova.  For example, did compact-object binaries with low-mass companions always have a
low-mass companion, or were they initially intermediate-mass X-ray binaries with the companion losing the bulk of its mass since formation \citep{Podsiadlowski:2004,Fragos:2015}?  Did high-mass X-ray binaries such as Cygnus X-1 experience previous mass transfer from the current optical companion \citep{Neijssel:2020CygX1,Zepei:2024}?  

Several authors have attempted to use either populations of binaries \citep[e.g.,][]{Richards:2022,Kapil:2022} or individual systems \citep[e.g.,][]{Willems:2005,Fragos:2009,Wong:2012} to place constraints on natal kicks by comparing observed properties against population synthesis models.  However, as long as large uncertainties persist in our understanding of other aspects of binary evolution, such as the common-envelope phase, such comparisons are inevitably sensitive to systematic biases.

Non-interacting binaries are perhaps the most promising in this regard.  If tidal interactions can be safely neglected, the present-day systemic velocity of the binary and its eccentricity are jointly sufficient to resolve the degeneracy between mass loss (Blaauw kick) and natal kick.  For example, the detached BH binary VFTS 243 \citep{Shenar:2022} allowed \citet{VignaGomez:2024} to place upper limits of $\lesssim 0.6$ M$_\odot$ on the ejected mass and $\lesssim 10$ km s$^{-1}$ on the natal kick. Along with Cygnus X-1, VFTS 243 strongly indicates that at least some BHs form via complete mass fallback without an appreciable natal kick.

\subsection{Compact-object binaries: rates and orbital properties}

As noted above, the very existence of binaries containing compact objects constrains the natal kicks to not be so large that they would disrupt the binary.  Population synthesis models of binary evolution frequently consider the impacts of natal kicks on the survival of compact-object binaries and on their orbital properties.  The literature is very rich here, but in addition to the results already mentioned in this review, some of the examples of studies that specifically considered the impacts of varying natal kicks on populations of X-ray binaries, double NSs, and gravitational wave sources include \citet{BrandtPodsiadlowski:1994,FryerKalogera:1997,Lipunov:1997,Postnov:2008,BeniaminiPiran:2016,Tauris:2017,Wysocki:2017,VignaGomez:2018,BrayEldridge:2018,GiacobboMapelli:2018,Broekgaarden:2022}.  

For example, when it comes to the formation of double compact objects, such as Galactic double NSs or gravitational-wave sources, most of these studies find that the first-born compact object must form when the binary is still wide.  It must therefore receive a low natal kick, smaller than the orbital velocity of the binary, in order for the binary to remain bound.  Subsequent mass transfer onto this compact object from the post-main sequence secondary, perhaps via a common-envelope phase, can then harden the binary, shrinking the separation by one to three orders of magnitude.  The kick of the secondary may therefore be larger without endangering binary survival.  Yet at least some Galactic double NSs exhibit small eccentricities $e \lesssim 0.1$, constraining the secondary's natal kick magnitude and the amount of mass lost in its supernova to moderate values, perhaps because the secondary is ultra-stripped by mass transfer from a helium giant progenitor prior to supernova \citep{Tauris:2015}.  On the other hand, the apparent bimodality in the eccentricity distribution of short-period Galactic double NSs suggests that at least some of the second supernovae do involve significant kicks and/or mass loss, or perhaps that an entirely different formation channel is at play \citep{AndrewsMandel:2019}.  Natal kicks may also lead to ultra-fast mergers \citep{BeniaminiPiran:2024,VignaGomez:2025}, possibly with associated spin flips (see below).

Unfortunately, there are a great many other uncertain parameters besides kicks that impact the rates at which binaries of a given type form.  Moreover, in many cases, even the fundamental channel of binary formation is uncertain.  For example, gravitational-wave sources may be forming through a combination of isolated binary evolution, dynamical formation in dense stellar environments, in hierarchical triples, or in the disks of active galactic nuclei \citep{Mapelli:2021, MandelFarmer:2018}.   Although most of these channels are sensitive to natal kicks, the impact of kicks on rates and orbital properties is sensitive to both the channel and to the detailed assumptions within that channel \citep{MandelBroekgaarden:2021}.  Great caution is therefore warranted before we can claim to be able to disentangle natal kicks from other uncertain physics.

\subsection{Spin/orbit misalignments}

Somewhat more promising than the binary rates and orbital properties are the angles between compact-object spins and the orbital angular momentum in binaries.  At least for tight isolated binaries, one may expect that the rotation of progenitors of compact objects is aligned to the orbit by a combination of tidal locking and mass accretion.  Therefore, if the spin axis of the compact object follows the spin of the progenitor (a strong assumption which may not hold -- see Section \ref{s:models} for a discussion of models), a misalignment would point to a natal kick at the birth of the compact object that changed the orbital plane \citep{Kalogera:2000}.

The population of gravitational-wave sources has been argued to show evidence of precession, pointing to spin-orbit misalignment and therefore, in the framework above, to significant natal kicks with a magnitude comparable to the pre-supernova orbital velocity of the binary \citep{GWTC2:pop,GWTC3:pop}.  However, precession is challenging to observe in short-lived binary BH mergers; for example, one event, GW200129 \citep{Hannam:2021}, may have been impacted by a data issue \citep{Payne:2022}, while another, GW190521 \citep{GW190521:astro}, could be explained by an eccentric binary \citep{RomeroShaw:2020GW190521} rather than a precessing one \citep{Miller:2024}.  GW200115, an NS-BH merger found to have evidence of misalignment \citep{GW200105}, is consistent with no evidence for spin at all \citep{MandelSmith:2021}.  In fact, \citet{Hoy:2024} find that only one of the first $\sim 70$ binary BH mergers detected via gravitational waves showed convincing evidence of precession.  

Meanwhile, the second pulsar in the double pulsar system PSR J0737-3039B is misaligned from the orbital angular momentum by 130 degrees \citep{Farr:2011}.  The low eccentricity of the binary is inconsistent with a very large kick that could have flipped the orbit; therefore, presumably, the spin is not aligned with the orbit, either because the progenitor avoided alignment or because the angular momentum direction was not conserved during the supernova.  The B pulsar spin is quite low, with a spin period of 2.8 s, which could make it easier to flip the angular momentum due to stochastic processes driving angular momentum ejection during a supernova  \citep{BlondinMezzacappa:2007}.

The BH in the low-mass X-ray binary MAXI J1820+070 has a spin that is misaligned from the orbital angular momentum by at least 40 degrees according to polarimetric observations \citep{Poutanen:2021}.  Although electromagnetic BH spin measurements are notoriously challenging \citep{MillerMiller:2015}, the presence of a jet in this system could indicate that the BH is at least moderately rapidly spinning.  The spin of the BH is unlikely to have changed significantly since birth, so this system could provide stronger evidence for a significant kick that tilted the binary's orbit \citep{Fragos:2010}.

However, as for possibly misaligned gravitational-wave sources, there remains a possibility that systems such as MAXI J1820+070 were formed through dynamical interactions.  Moreover, as discussed above, compact objects may not retain the spin direction of their progenitors \citep[see also][]{Larsen:2025}.  If so, the misalignment does not necessarily indicate a natal kick.

\subsection{Merger locations}

Short gamma-ray bursts, as well as some long gamma-ray bursts \citep{Rastinejad:2022,Levan:2023}, are associated with mergers of pairs of NSs \citep{GW170817:MMA} or NS-BH mergers.  Short gamma-ray bursts appear to be more broadly spatially distributed within their host galaxies than long gamma-ray bursts associated with collapsars \citep{FongBerger:2013, 2022ApJ...940...56F, 2022MNRAS.515.4890O}, although \citet{PeretsBeniamini:2021} suggest that this may be an artifact due to differences between typical host galaxy morphologies.  Moreover, some burst locations do not coincide with any known galaxies.  It is natural to assume that natal kicks are responsible for the spread of merger sites within the hosts away from star-forming regions and for the ejection of merging binaries from host galaxies,  see e.g., \citet{2025A&A...699A.113G} and references therein. 

Unfortunately, the sample of well-localized bursts likely suffers from significant selection effects.  The accurate localization of bursts requires the afterglow to be observed, but the luminosity, duration, and detectability of an afterglow depend on the density of the surrounding medium \citep{Berger:2010}, though see \citet{Tunnicliffe:2013}.  Moreover, the association of apparently hostless bursts with a particular host galaxy is uncertain, and even if such an association is accurate, the distance is only known in projection.  

In principle, if NS mergers are responsible for the production of third-peak r-process elements such as europium, the distribution of such elements within our Galaxy could provide information about natal kicks.  However, the yield of r-process elements from a merger, the rate of mergers, the amount of turbulent mixing, and the contribution of other processes such as magneto-rotational hypernovae to heavy r-process nucleosynthesis are too uncertain at present for this to be a useful constraint on kicks \citep[e.g.,][]{Kobayashi:2023}.  On the other hand, the presence of heavy r-process material in ultra-faint dwarf galaxies under the assumption that NS mergers are responsible for its production supports models in which at least some NSs receive low kicks, allowing NS binaries to be retained and merge promptly \citep{Beniamini:2016}.

\section{Kick magnitude phenomenological models}

Population synthesis studies of massive stars rely on natal kick models as a key ingredient in the forward modelling of stellar and binary evolution, leading to the formation of transients such as gravitational-wave sources, interacting supernovae, and kilonovae.  Natal kicks are also an important ingredient for modelling the initial population of compact objects in dynamical models of dense stellar environments such as globular clusters.  Typically, such models rely on one of two approaches: either a single distribution from which natal kicks are drawn for compact objects or natal kick models based on the properties of progenitors at the time of explosion, although some mixture of these is occasionally adopted.  We will briefly summarize both approaches below.

\subsection{Population-level distributions}

A single population-level distribution is often the default approach for modelling NS kicks.  The distributions summarized in Section~\ref{s:pulsars} are frequently applied as NS natal kick distributions, even though these typically describe the observed velocity distributions of single pulsars, not the initial kicks.  However, this may not be as problematic as one might imagine.  Although NSs are typically born in binaries, the ones receiving kicks significantly larger than the orbital velocity escape with velocities that are similar to the natal kick, only weakly perturbed by the orbital velocity. The ones receiving kicks significantly smaller than the orbital velocity do not escape at all unless the binary is unbound by mass loss, and do not enter the single-pulsar population (though this does place a selection effect against the weakest kicks in the single-pulsar observations). Only NSs with kicks comparable to the orbital velocity enter the single-pulsar population with velocities that may be quite different from their natal kick velocity, but these are a minority \citep{Kapil:2022}.

In general, most of the models for NS natal kicks described in \ref{s:pulsars} represent plausible choices, though the single-Maxwellian model of \cite{2005MNRAS.360..974H} should be corrected to use $\sigma = 217$ km~s$^{-1}$ to account for their missing Jacobian \citep{2025arXiv250522102D}.

Some modellers single out specific NS formation scenarios to receive reduced kicks.  Electron-capture supernovae are often assumed to impart reduced kicks of only a few tens or even a few km~s$^{-1}$ \citep{Pfahl:2002let,Podsiadlowski:2004,Gessner:2018}.   NSs formed through the accretion induced collapse of oxygen-neon white dwarfs are often placed in the same category.  Similarly low kicks are sometimes applied to the products of ultra-stripped supernovae: NSs formed in tight binaries (often with other compact object companions) from naked helium star progenitors that expanded and lost their helium envelopes during case BB mass transfer and emerged with carbon-oxygen cores with just enough mass to collapse into a compact object \citep{Tauris:2015}.   

Given the paucity of direct observational evidence for BH kicks, 
these are sometimes selected from a distribution similar to NS kicks, but down-weighted according to the fraction of mass falling back onto the proto-NS during the formation of the BH $f_\mathrm{fb}$, so that the NS kick is scaled by $(1-f_\mathrm{fb})$ to obtain the BH natal kick \citep{Fryer:2012}.  BHs forming through the complete collapse of the progenitor are often assumed not to receive a natal kick.  However, other choices, such as the assumptions that BHs receive the same {\it momentum} kick as NSs, with the BH natal kick then scaled down by the BH-to-NS mass ratio, also appear in population synthesis modelling.

The choice of kick direction is another uncertain parameter that has important consequences for binary evolution.  Although isotropic kicks are standard, some authors have considered the impact of kicks that may be preferentially oriented either in the equatorial plane of the rotating progenitor (which is often nearly aligned with the pre-SN orbital plane by the action of mass transfer and tides) or, conversely, in the polar direction.

The consequences of post-natal rocket kicks on binary evolution are just beginning to be explored in detail.  While these may not be sufficient to explain the full observed natal kicks, they could play an important role in some classes of binaries, such as wide NS binaries \citep{2024ApJ...972L..18H,2025arXiv250705602B}.

\subsection{Models for kicks based on progenitor properties}

The assumption that an NS natal kick is drawn from a single distribution runs counter to the intuition that the kick is set by the physics of the explosion and, therefore, should be sensitive to the state of the progenitor at the time of the explosion.  In fact, the variations for electron-capture and ultra-stripped supernovae discussed above, as well as the dependence of BH natal kicks on the remnant mass or fallback fraction, already attempt to capture this relationship.  However, the apparent correlation between NS mass and kick observed in NS-forming core collapse supernova simulations (see Section~\ref{s:models}) calls for a kick model that would naturally allow for such correlations.

\cite{BrayEldridge:2016, BrayEldridge:2018} and \cite{GiacobboMapelli:2020} proposed models inspired by the conservation of momentum during the supernova, with the kick velocity a linear function of the ratio of the ejecta mass to the NS mass.  However, these linear models considered the entire mass of the ejecta. Moreover, the fitted model parameters allowed for seemingly unphysical negative kick speeds in some cases \citep{BrayEldridge:2016},  though see the recalibrated values from \citet{Richards:2022}. 

\citet{MandelMueller:2020} also assumed that momentum must be conserved.  However, on the basis of supernova simulations, they posited that material far from the core is ejected promptly before appreciable asymmetries build up, and only material from the carbon-oxygen core can attain a significant degree of asymmetry.  Thus, their remnant kicks scale as 
\begin{equation}
v_\mathrm{kick} = \left( v_\mathrm{NS/BH} \frac{M_\mathrm{CO} - M_\mathrm{remnant}}{M_\mathrm{remnant}} \right) (1+\sigma_\mathrm{NS/BH}),
\end{equation}
where $M_\mathrm{CO}$ is the progenitor's carbon-oxygen core mass, $v_\mathrm{NS}$ or $v_\mathrm{BH}$ are parameters describing the combination of ejecta velocity and asymmetry which are assumed to be universal for NSs or BHs, respectively, but must be calibrated to models or observations, and $\sigma$ is a dimensionless scatter parameter that allows for stochasticity in kicks.

\citet{Kapil:2022} calibrated the free parameters related to NS kicks by matching the outputs of single-star and binary population synthesis to single-pulsar velocity observations.  They found that $v_\mathrm{NS} = 520$ km~s$^{-1}$ and $\sigma_\mathrm{NS}$=0.3 represent a good fit to the data. 

Instead of relying on a simple parametric prescription, \citet{VignaGomez:2018} used a relatively complex fit for the kicks obtained as an extension of the semi-analytic supernova model of \citet{Mueller:2016}. The downside of tying the kick prescription to actual predictions of supernova explosion properties is that such a prescription becomes very dependent on the underlying set of stellar evolution models. It may therefore include features that are not universal and robust outcomes of stellar evolution models.

\section{Conclusions and open questions}

In this review, we have comprehensively discussed the various observations that demonstrate that most NSs and a significant fraction of BHs obtain an additional velocity kick at birth. For NSs, these kicks can be $\gtrsim1000$~km~s$^{-1}$ and for BHs $\gtrsim100$~km~s$^{-1}$. Large kicks of NSs and  all BH kicks are inevitably related to SN explosions. Kicks play an essential role in binary evolution and in the dynamics of compact objects. 

Still, there is a plethora of unsolved problems related to the kick velocities of compact objects. In the first place, the shape of the initial velocity distribution is unknown. As there are several possible mechanisms (not all of them directly related to an SN explosion) for a compact object to acquire an additional velocity at birth, the distribution is expected to be multi-component, especially for NSs. Even if some specific cases are relatively rare in comparison to the total number of compact objects, they can be of crucial importance for some types of objects. 

Progress in instrumentation (Grace Roman Space Telescope, SKA, and other new facilities) will lead to a huge increase in the number of measured spatial velocities (mostly their sky projections) of compact objects in the near future. However, most observations do not directly probe low-velocity kicks and kicks in surviving binary systems. 

The fraction of low-velocity compact objects is one of the most important ingredients in modeling NSs in globular clusters, wide binary systems (like those recently discovered by Gaia), and for hypothetical isolated accreting NSs. 

There are several strong claims in favour of a bimodal or multimodal kick velocity distribution for NSs. If this is indeed the case, it is essential to understand the origin of this multimodality. If the modes arise due to variation in the SN explosion mechanism or different progenitor evolution channels, then it is necessary to explore if NSs comprising different modes of the distribution have systematically different parameters (masses, initial spin periods, initial magnetic fields, etc.). This task is non-trivial, as most physical parameters of NSs (especially their initial values) are not measured in a model-independent way. 

BH kicks may also demonstrate a kind of bimodality. Nearly zero kicks may arise in the case of a direct collapse. By contrast, significant kicks are expected during delayed BH formation due to fallback onto a proto-NS.  

The list of issues for future studies can be continued: the role of binary (triple) evolution in the value of the kick, spin-velocity correlation, preferential kick direction, etc. Uncertainties in all these properties significantly limit our ability to perform precise modelling of populations of various source types with NSs and/or BHs. We look forward to significant advances in our understanding of the properties and origin of the kick velocity of compact objects in the near future. 

\section*{Acknowledgements}

The authors thank Paul Disberg, Ryosuke Hirai, and Steinn Sigurdsson for their insights, and especially Andrei Igoshev for many discussions and comments at the early stage of the project. We also thank the referee for useful comments and suggestions. IM acknowledges support from the Australian Research Council (ARC) Centre of Excellence for Gravitational Wave Discovery (OzGrav), through project number CE230100016. The work of IM was performed in part at the Aspen Center for Physics, which is supported by National Science Foundation grant PHY-2210452.

\appendix




\bibliographystyle{elsarticle-harv} 
\bibliography{references}






\end{document}